\documentclass[10pt,final,twocolumn,twoside]{IEEEtran}
\usepackage{graphicx} \usepackage{amsmath} \usepackage{amssymb} \usepackage{amsthm} \usepackage{algorithm} \usepackage{algpseudocode} 
\usepackage{cite}
\usepackage[caption=false,font=footnotesize]{subfig}
\usepackage{stfloats}
\usepackage{url}
\usepackage{mathtools}

\usepackage{multirow}
\usepackage{color}
\usepackage{tablefootnote}
\usepackage{tikz}
\usepackage{relsize}

\usetikzlibrary{decorations.pathreplacing}

      \theoremstyle{definition}      \theoremstyle{remark}      \theoremstyle{plain}     

           \newcommand{\mbf}[1]{\mathbf{#1}}     \newcommand{\mbs}[1]{\boldsymbol{#1}}

\newcommand{\TD}{T_{\scriptsize \mbox{D}}}
\newcommand{\TR}{T_{\scriptsize \mbox{R}}}
\newcommand{\mb}[1]{\mbox{#1}}

\newcommand{\xunit}{0.00085\columnwidth}
\newcommand{\yunit}{0.12\columnwidth}

\newcommand{\MS}[1]{\scriptsize \mbox{#1}}
\newcommand{\suchthat}{\, \mid \,} 
\begin{document}

\title{ Optimal Scheduling for Interference Mitigation by Range Information}
\author{Vijaya Yajnanarayana, Klas E. G. Magnusson, Rasmus Brandt, Satyam Dwivedi, Peter H\"{a}ndel
\thanks{

    $\bullet$ V. Yajnanarayana, K. Magnusson, S. Dwivedi and P. H\"{a}ndel are with Department of Signal Processing, KTH Royal Institute of Technology, SE-100 44 Stockholm, Sweden. (email\{vpy,klasma,ph\}@kth.se)

$\bullet$ R. Brandt is with Google research, Stockholm, Sweden. (email: rabr5411@kth.se)
}
}

\maketitle

\begin{abstract}
The multiple access scheduling decides how the channel is shared among the nodes in the network. Typical scheduling algorithms aims at increasing the channel utilization and thereby throughput of the network. This paper describes several algorithms for generating an optimal schedule in terms of channel utilization for multiple access by utilizing range information in a fully connected network. We also provide detailed analysis for the proposed  algorithms performance in terms of their complexity, convergence, and effect of non-idealities in the network. The performance of the proposed schemes are compared with non-aided methods to quantify the benefits of using the range information in the communication. The proposed methods  have several favorable properties for the scalable systems. We show that the proposed techniques  yields better channel utilization and throughput as the number of nodes in the network increases.  We provide simulation results in support of this claim. The proposed methods indicate that the throughput can be increased on average by $\mbf{3-10}$ times for typical network configurations. 
\end{abstract}

\begin{IEEEkeywords}
Sensor networks, ad-hoc networks, mobile networks, swarm networks, cooperative communication, position dependent communication, ultra wideband (UWB) communication, mmWave communication, 5G-communication, traveling sales man (TSP) problem.
\end{IEEEkeywords}
\vspace{-0.15in}
\section{Introduction}
\label{sec:intro}

The recent advances in sensor technology have resulted in development of low-cost, low-power sensors, which are capable of sensing, data processing, and communication. Many sensor networks have a  large number of sensor nodes, which are densely deployed over a wide geographical region to track a certain physical phenomenon \cite{Hill-sensorplatform,rhee-sync}. These sensors could have a fixed topology, as in the case of smart sensors used in structural health monitoring \cite{NSEL} or have a dynamic topology, as in the case of sensors mounted on autonomous robots for applications discussed in \cite{AndersanSwarm,MondadaSwarm,VijayKumar}.

\begin{figure} [!t]
  \centering
\includegraphics[width=2.5in]{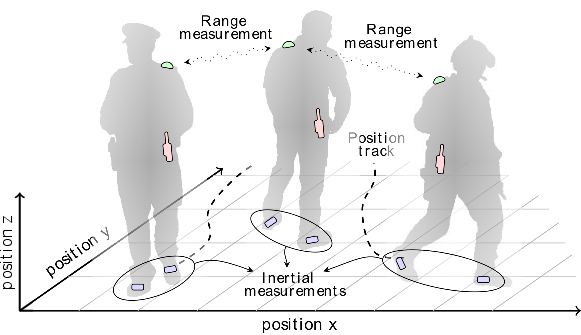}
  \caption{Illustration of fire fighters agents sharing information continuously with other agents  \cite{JO-interagent}.}
  \label{fig:firefighter}
\end{figure}

In sensor networks, there are many situations where every node needs to transmit a message to every other node at regular intervals. This type of communication is typically required for information dissemination across the network to accomplish various tasks such as localization, routing, distributed control and computation. For example, in \cite{JO-interagent,Rantakokko-2} firefighter agents share information at regular intervals through point to multi-point communication, where every agent broadcasts  sensor data, like position, temperature, visibility, etc., to all other agents. This enables every firefighter to know relevant information about other firefighters, thereby increasing the efficiency of operation. This is illustrated in Fig.~\ref{fig:firefighter}. This type of communication can also be found in the cooperating swarm of micro unmanned aerial vehicles (UAVs). These are low payload carrying, scaled down quadrotor platforms with relevant sensors mounted on them \cite{VijayKumar}. Constant updates (communication) between sensors are essential in many UAV networks, as they need to coordinate to accomplish the required tasks. These updates could include sensor data, position information, etc. Similar regular broadcast communication by sensor nodes can also be found in other swarm networks as discussed in \cite{AndersanSwarm,MondadaSwarm}. Reporting the health of the sensor node to all other nodes in wireless sensor networks (WSN) described in \cite{Bharathidasan} requires regular communication by the sensor nodes. In underwater acoustic (UWA) sensor networks broadcast communication of similar nature is used for time-synchronization, coordination, self-configuration and localization \cite{klen-sched,Akyildiz-UWA}. All these networks employ some form of all-to-all broadcast communication between nodes.

Sensor networks in which each sensor has to share information constantly with the other sensors through all-to-all broadcast can be accomplished efficiently by communicating through a shared broadcast channel.  As the density of the sensor network increases, the effective bitrate per sensor, $R_s$, drops, since the total bitrate, $R_b$, supported by the shared broadcast channel is fixed. In many sensor networks, there exists a long propagation delay in communication in relation to the scheduled access duration (packet length)\footnote{Packet length and access duration are used interchangeably.} of the shared channel. This can arise either due to low propagation speed of the physical layer signal in the medium or large distances between sensors (geometric size of the topology). For example,  in UWA sensor networks, the  propagation of acoustic physical layer signal in water is five orders of magnitude slower than in wireless radio channel, coupled with the large distances between sensors in oceans make the above scenario common in these networks. Similar scenarios exist in few wireless sensor networks (WSN) employing impules radio UWB (IR-UWB) and millimeter wave (mmWave) technologies. In IR-UWB and mmWave channels  with high directivity gain  can have delay spread of order of few tens of nano-seconds, thus can have small access schedules \cite{Karedal-UWB-Industrial,Zahedi-Channel,Rappaport-directionality}.

Our intention in this paper is to develop an efficient broadcast schedule to access the shared channel for the sensor network by exploiting the propagation delays between sensor nodes. We define  one report cycle (update cycle), transmitting $\TR$, as the total time duration during which all the nodes in the sensor network have transmitted and received one message packet to and from all the other nodes in the network. We use this performance metric to assess performance of various schemes proposed in the paper.

\subsection{Related Work}
\label{ss:related}

The main aim of the paper is to propose methods that optimize the multiple access schedule by exploiting the spatial-temporal aspect of the channel for the problem discussed in Section \ref{sec:intro}.  As will be shown in Section \ref{sec:sys-model}, this can be posed as an optimization problem, the solution for which is non-convex and computational complexity scales exponentially with the increase in the number of nodes. The are several works in UWA networks, where this problem is addressed, particularly for accomplishing  tasks such as self localization. For example, in \cite{Park-AllToAll}, the interference free all-to-all broadcast in the UWA sensor networks is posed as an optimization problem, which is similar to the problem formulation in Section  \ref{sec:sys-model}. A suboptimal solution is obtained in \cite{Park-AllToAll} using a heuristic method, which relaxes the constraints to enable schedule computation for the nodes in a sequential order. However, a better schedule can be obtained, by increasing the computational complexity by changing  the optimization problem, so that it can be solved using convex methods, traveling salesman problem (TSP) and iterative path-adjusting methods proposed in this paper. This is discussed further in Section \ref{sec:ALG}. Broadcast schedule construction for a partially connected network for localization tasks is discussed in \cite{Ramezani}, which allows transmissions on distinct sub channels. A special case of this formulation with a single channel construction can be considered for solving the problem considered in this paper. However, to ensure that the acoustic signals in UWA networks do not collide in the space between anchors nodes (where possible sensors-nodes may exist) more stringent constraints are enforced. This may not be applicable for electromagnetic wave based WSN networks as there is no separation between the anchor-node and ordinary sensor-node in the communication and only interference at the receive node needs to be nulled. This is further discussed in Section \ref{subsec:CVX}. Chen \emph{et al., }\cite{Chen-CSMA} discuss the broadcast schedule construction for a UWA network using the traveling salesman approach. This is similar to the TSP problem discussed in this paper. However, in \cite{Chen-CSMA}, UWA anchor-nodes cannot transmit before it has received the message from the previous node in the sequence to avoid collision of acoustic waves at the sensor nodes. This will result in a symmetric TSP problem. In the context of a general all-to-all communication in WSN this constraint is not needed and this will manifest the problem as an asymmetric TSP problem as shown in Section \ref{subsec:tsp}. All the posposed methods discussed above (including the ones discussed in this paper), require a centralized sensor network, with a powerful coordinator node, which exploits the position information from all participating nodes for all-to-all broadcast schedule construction. In the networks, where the centralized configuration is not possible  or position information of the nodes is unavailable, the algorithms proposed in \cite{Ramezani-Collision,Ferrari2011} can be employed, however, as will be shown in Section \ref{sec:SS}, solving the broadcast schedule optimization problem using position information can significantly improve performance.

In the work proposed in  \cite{Nico-Broadcast}, an adaptive push system for information dissemination is designed. Here the broadcast schedule is created at the server for a network with varying client node demands using learning automata (LA).  Similar scheduled communication on a broadcast channel for dedicated traffic flow is discussed in \cite{klen-sched,Klen-becaon}.  In contrast to these papers, we are interested in all-to-all communication with fair access\footnote{Fair access here indicates that within one report cycle, $\TR$ all the nodes in the network will get one access to the shared common channel for  transmitting one packet.} to all the nodes in the network to the broadcast channel as discussed in  Section \ref{sec:intro}. 

There also exists standard time division multiple access (TDMA) schemes such as  slotted floor acquisition multiple access (FAMA) where regulated transmissions for all the nodes can be accomplished \cite{Slotted-FAMA}. However, in a regular time division channel, the shared common channel is slotted in time and each one of the $N$ nodes of the sensor network will have access to a time slot which is a uniform fraction of the report cycle, $\TR$. As shown in Section \ref{sec:sys-model}, as the radius of the sensor network topology and the number of nodes in it increase, the throughput per sensor and the update rate decrease. By exploiting the range information, orthogonality can still be maintained for overlapping time slots which leads to higher capacity. For ideal positioning of nodes, the throughput can be increased by $N$ times, leading to a significant performance gain. Even when the positions of nodes are randomly distributed, the performance boost can be substantial in practice. For the realistic examples studied in this paper, the throughput is increased by an order of magnitude ($10$ times for $100$ node configurations with outliers as discussed in Section \ref{sec:SS}) compared with a regular scheme. 

The main contributions of this paper are as summarized below.
\begin{itemize}
\item We introduce three novel methods, which exploit the range information for efficient communication for the broadcast problem discussed.
\item We analyze these methods in terms of computational complexity 
\item We discuss the performance analysis of these methods for   different topologies and contrast them with standard multiple access protocols such as code division multiple access (CDMA).
\item We discuss the sensitivity of these protocols to the non-idealities such as range and synchronization errors.
\end{itemize}

\begin{figure} [!t]
  \centering
  \includegraphics[scale=0.4]{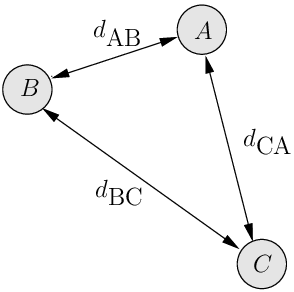}
  \caption{Peer-to-peer ad-hoc sensor network with $3$ nodes. $d_{\mbox{AB}}$, $d_{\mbox{BC}}$ and $d_{\mbox{CA}}$ are the path lengths between nodes $A$, $B$ and $C$.}
  \label{fig:p2p}
\end{figure}

We will demonstrate the methods using a simple $3$ node network shown in Fig. \ref{fig:p2p}. This will aid us in explaining the algorithms clearly. Subsequently, we will demonstrate the performance of the proposed methods in different network topologies of varied sizes. We use the report cycle, $\TR$, as a metric to assess performance, with the objective to minimize this parameter. 

The rest of the paper is organized as follows: In Section \ref{sec:sys-model}, we discuss the system model and formulate the problem. In Section \ref{sec:ALG}, we propose algorithms which exploit the range information to provide efficient communication between nodes.  In Section \ref{sec:sync}, we study the effect of synchronization and range errors on the proposed algorithms. In Section \ref{sec:cdma}, we compare the effective bitrate per sensor, $R_s$, of the proposed algorithms with the code division multiple access (CDMA) approach. In Section \ref{sec:SS}, we evaluate the proposed methods for a large number of nodes with different topologies and demonstrate the performance gain of utilizing the range information. Finally, in Section \ref{sec:con} we discuss the conclusions.

\section{System Model and Problem Formulation}
\label {sec:sys-model}


Consider a general setup of a fully connected sensor network with $N$ nodes. For the sake of the discussion, we set  the access duration (message packet length)  per node to be $\tau=100$ time units. We define the path equivalent message length as $\mathcal{L}=\mu \tau$ length units, where  $\mu$ is the velocity of the physical layer signal in the propagation medium. The message packets are said to be correctly received, if the packets do not interfere, i.e., there is no collision of packets at the receiving node.


\begin{table} [!t]
  \centering
  \caption{Configuration for the topology in Fig. \ref{fig:p2p}.}
  \begin{tabular}{|c|c|}
    \hline
    \textbf{Parameter} & \textbf{Value}\\
    \hline
    $d_{\MS{AB}}$ & $95~\mb{m}$\\
    $d_{\MS{BC}}$ & $110~\mb{m}$\\
    $d_{\MS{CA}}$ & $105~\mb{m}$\\
    $\mathcal{L}$ & $30~\mb{m}$\\
    $\tau$ & $100~~\mb{ns}$ \\
    $\mu$ & $3\times10^8~\mb{m/s}$\\
    \hline
  \end{tabular}
  \label{tab:cfg}
\end{table}

\begin{figure*}[!t]
  \centering
  \subfloat[Interfering message packets due to concurrent transmission.]{\includegraphics[width=8.8cm]{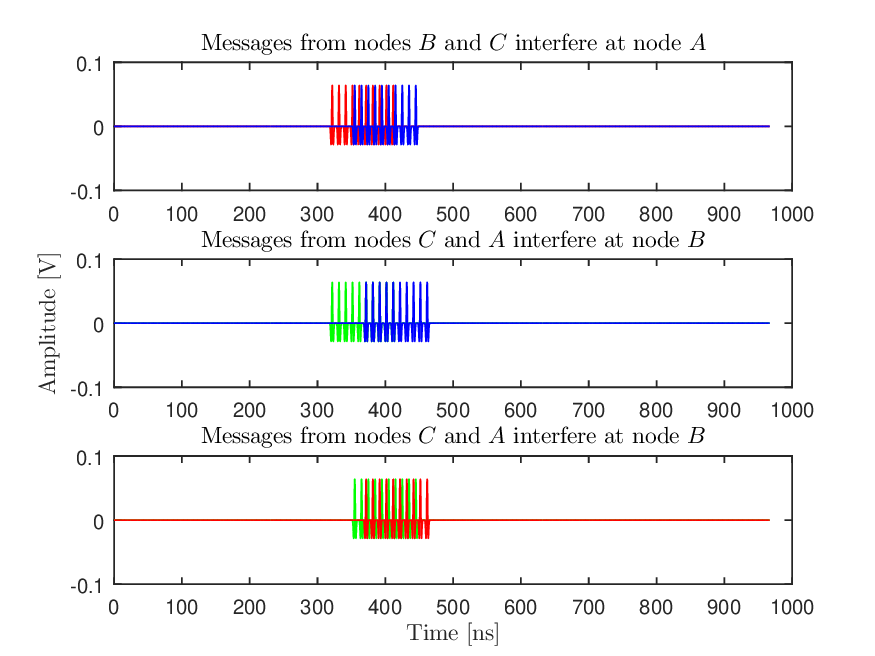}         \label{fig:inter}}
\subfloat[Arrival of packets without interference at $A$, $B$ and $C$ nodes after introducing delays of $\Delta_{\MS{B}}=84\,\MS{ns}$ and $\Delta_{\MS{C}}=150\,\MS{ns}$ in B and C nodes, respectively.]{\includegraphics[width=9cm]{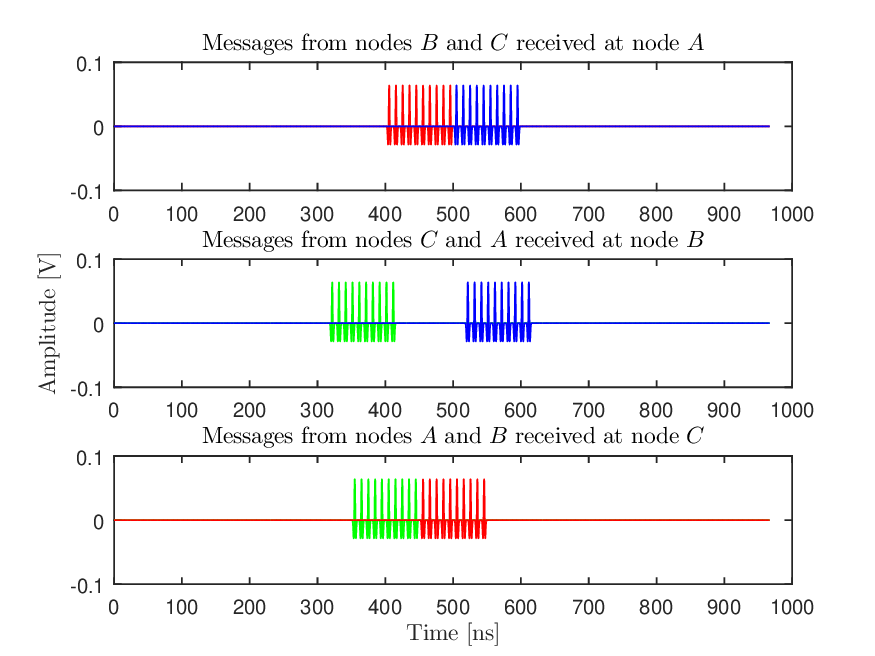}
  \label{fig:clean}}
\caption{Concurrent transmission on shared common channel will result in interference as shown in (a). If we solve the optimization problem defined in \eqref{eq:opt} then the interference can be mitigated as shown in (b). The signal representing the message packet from nodes $A$, $B$ and $C$, $\left(p_i(t)\suchthat i \in \{A,B,C\}\right)$ is shown in green, red, and blue respectively. The $\tau=100~\mb{ns}$, $\mu=3\times10^8~\mb{m/s}$, and $\mathcal{L}=30 ~\mb{m}$ is considered in the illustration. }
\label{fig:train}
\end{figure*}

\subsection{Orthogonalization with scheduled transmission}
\label{subsec:orthogonal}
 In a network of $N$ nodes, if we assume that the $K =$$ N\choose 2$ range values are available, one approach to orthogonalize the transmission is by creating a sequential schedule, where each node gets to transmit a message every $\TD$ time units, where $\TD$ is given by
\begin{equation}
  \label{eq:td}
  \TD=\frac{D}{\mu} + \tau,
\end{equation}
where $D$ is the maximum of the $K$ range values, that is
\begin{equation}
  \label{eq:D}
  D=\underset{i,j}\max\{d_{ij}\},\forall i,j\in [1,\ldots,N],i\ne j.
\end{equation}
With this approach, one report cycle, $\TR$ is given by

\begin{equation}
  \label{eq:TR}
  \TR=N\TD.
\end{equation}

To exemplify the above discussion, we consider a $3$ node peer-to-peer network as shown in Fig~\ref{fig:p2p}. For the sake of discussion, the nodes are labeled as $A$, $B$, and $C$. From \eqref{eq:td} and \eqref{eq:D}, we get
\begin{equation}
  \label{eq:td3}
  \TD=\frac{\max\{d_{\MS{AB}},d_{\MS{BC}},d_{\MS{CA}}\}}{\mu}+\tau=\frac{d_{\MS{BC}}}{\mu}+\tau.
\end{equation}
From \eqref{eq:TR}, notice that the report cycle, $\TR$, increases linearly with the number of nodes in the network ($N$) and the radius of the network topology ($D$). Therefore, as the number of nodes  or the geometric size of the network increases, $\TR$, will increase, resulting in inefficient utilization of the shared common channel; thus, requiring an improved communication method.

In many  networks, the geometry of the sensor placements is such that the difference in propagation time for the message packets to arrive at nodes are larger than the duration of the message packets themselves. These situations arise in many sensor networks which have small message packets to be shared with other sensors, resulting in a very small value of $\mathcal{L}$. This situation could also arise in future 5G networks, where the physical layer packet lengths of devices in a macro cell  are much smaller (on the order of a few microseconds) compared to the cell dimensions (on the order of a few kilometers)$\mbox{\cite{Rappaport-book,Rappaport-5G, Agyapong-5G, Panzner-5G}}$. We can reduce the report cycle of the network by exploiting this fact. Consider a sensor network in which the path difference between any two nodes is greater than $\mathcal{L}$. Then, concurrent transmissions will result in message packets arriving at different times at each node, hence all transmissions are orthogonal.  In general, for an $N$ node network to ensure concurrent orthogonal transmissions, the network should fulfill the conditions
\begin{equation}
  \label{eq:con}
\begin{aligned}
  & |d_{ki}-d_{kj}| \ge \mathcal{L} \\
  & \forall\mb{ } i,j,k \in [1,2,\ldots,N] \suchthat i,j \ne k \mb{ and } i\ne j,\\
\end{aligned}
\end{equation}
where $i,j$ and $k$ denote the distinct nodes in the network
and $d_{ki}$ and $d_{kj}$ denote the distance from the $k$-th node to node $i$ and node $j$ respectively. Thus, the report cycle, $\TR$, is equal to the maximum path delay, $\TD$, in the network, instead of $N\TD$ for scheduled transmission as discussed before.

For example, consider the $3$ node network shown in Fig.~\ref{fig:p2p}. Suppose, the dimensions of $d_{\MS{AB}}$, $d_{\MS{BC}}$, and $d_{\MS{CA}}$ does not follow the specifications of Table~\ref{tab:cfg} and if $|d_{\MS{AB}}-d_{\MS{AC}}| \ge \mathcal{L}$, then the signal transmitted simultaneously at nodes $B$ and $C$ will arrive at node $A$ at different times, and hence $A$ can correctly receive them. Similarly, $|d_{\MS{BA}}-d_{\MS{BC}}| \ge \mathcal{L}$ and $|d_{\MS{CA}}-d_{\MS{CB}}| \ge \mathcal{L}$ will ensure correct message packet reception at nodes $B$ and $C$  respectively. Thus, all the three nodes can concurrently transmit, and the report cycle can be completed in $\TD$.

In general sensor networks, \eqref{eq:con}  is rarely fulfilled. When a network with $N$ nodes has a particular geometric configuration, which does not meet condition  \eqref{eq:con}, we can reduce $\TR$ by introducing a delay  $\Delta_i$ to each node $i\in [1,2,\ldots,N]$. The $\Delta_i$s are adjusted such that the message packets do not interfere at the receiving nodes. The $\Delta_i$s form the time schedule during which the $i\mbox{-th}$ node needs to transmit. The optimal schedule is obtained by solving the following optimization problem.
\begin{equation}
\label{eq:opt}
\begin{aligned}
& \underset{\{\Delta_i\}}{\text{minimize }}\max_{i,k}
& & \Delta_i + \delta_{ki} \\
& \text{subject to}
& & J=J_1+J_2+\ldots+J_N=0,
\end{aligned}
\end{equation}
where
\begin{equation*}
\label{eq:opt1}
\begin{aligned}
&J_k=\Big|\int\underset{ij}{\sum}p_i\left(t-\delta_{ki}-\Delta_i\right)p_j\left(t-\delta_{kj}-\Delta_j\right)dt  \Big|\\
&\forall\mb{ } i,j,k \in [1,2,\ldots,N] \suchthat i,j \ne k \mb{ and } i\ne j.
\end{aligned}
\end{equation*}
Here, $p_i(t),i\in[1,2,\ldots,N]$, denotes the physical layer signal of the message packet, $J_k$ denotes the interference due to the received message packets at node $k$, and $J$ indicates the total interference in the system. The $\delta_{ki}$ represents the \mbox{path-delay} between the $k\mbox{-th}$ node and $i\mbox{-th}$ node and is given by $d_{ki}/\mu$. 
The report cycle with this approach is given by
\begin{equation}
  \label{eq:TR_delays}
  \TR=\max_{i,j}(\Delta_i+\delta_{ji})+\tau, \forall i,j \in [1,\ldots,N]\mb{ and } i\ne j.
\end{equation}
To illustrate the  solution of the optimization problem \eqref{eq:opt}, we  once again consider the $3$ node network shown in Fig.~\ref{fig:p2p}. The configuration defined in Table~\ref{tab:cfg} is used for path lengths. In Table \ref{tab:cfg}, the path differences between nodes do not meet the constraint defined in \eqref{eq:con}. That is, if all the nodes transmit simultaneously, they will interfere with each other. For example, if at time $t=0$, all the nodes $A$, $B$ and $C$ concurrently transmit their message packets, then the received signal  at nodes $A$, $B$ and $C$ are shown in Fig.~\ref{fig:inter}.

To accomplish short report cycle without interference in the example discussed above, the optimization \eqref{eq:opt} is solved using the grid search method with $\tau=100~\mb{ns}$, $\mu=3\times10^8~\mb{m/s}$, and $\mathcal{L}=30 ~\mb{m}$. In this method, we set $\Delta_{\MS{A}}=0$; assuming that all nodes are synchronized to node $A$, $J$ is computed by varying $\Delta_{\MS{B}}$ and $\Delta_{\MS{C}}$ over the interval $\left[0,\TD\right]$, where $\TD$ is given by \eqref{eq:td}. The solution for the optimization problem using the grid search method yields $\Delta_{\MS{B}}=84\,\mb{ns}$ and $\Delta_{\MS{C}}=150 \,\mb{ns}$. With these delays introduced in nodes $B$ and $C$, the signals are not interfering, as shown in Fig. \ref{fig:clean}. Node $C$, will transmit last after a delay of $150\,\mb{ns}$  and the resulting report cycle using \eqref{eq:TR_delays} is $620\,\mb{ns}$.

\subsection{System Aspects}
\label{subsec:sysasp}
Consider a centralized sensor network, with a powerful coordinator node, which broadcasts a beacon message with a time-stamp and the registration request. The ordinary nodes will respond with their location information after synchronizing their clock using the time stamp in the beacon\footnote{Ordinary nodes can use TDOA method to account for the transmission delay during synchronization.}. This communication can employ a conventional TDMA scheme on a control channel. The coordinator solves the optimization problem \eqref{eq:opt} using the range values of the participating nodes to prepare the broadcast schedules for the nodes. This information is encapsulated into a control packet and transmitted to all the participating nodes. Periodically the central node need to collect the  information from the participating nodes to resolve the optimization problem to cater to the change in topology due to the  node mobility or node failures in the network.  Note that the physical control channel on which the registration request and the broadcast schedules are communicated are different from the shared common channel used for all-to-all communication. Even with a powerful coordinator the solution of  \eqref{eq:opt} is not possible as the scale of the network grows. In the later sections, we will discuss how a practical solution for  \eqref{eq:opt} can be achieved.

Even though the nodes clocks are synchronized during the initialization process, the synchronization can be lost due to the clock drift, jitter etc.  In many sensor networks,  synchronization is  accomplished using a message passing technique as proposed in timing-sync (TSYNC) or  reference broadcast synchronization (RBS) protocols \cite{timing-sync,RBS}.  Network synchronization ensures that all the nodes in the network have the same time scale. We also assume that exact range information  is available. Recently, there has been  some work on estimation algorithms for joint ranging and synchronization. These  are proposed in \cite{Satyam-joint, Rajan-joint}. These algorithms can yield joint accuracy levels up to few centimeters for range and few nanoseconds for synchronization. We study the behavior of the proposed algorithms in the presence of range and synchronization errors in Section \ref{sec:sync}.

\section{Algorithms}
\label{sec:ALG}
Using the grid search method to solve \eqref{eq:opt} is costly, as the algorithm complexity, $O(q^N)$, increases exponentially  with the number of nodes in the network. Here, $q$ indicates the size of the quantized grid of interval $\left[0,\TD\right]$ used in the grid search. In this section, we propose three distinct methods to solve the above problem, each having benefits over the other depending on the network geometry, complexity, etc.

Consider the arrival of messages at node $k$ from nodes $i$ and $j$ as shown in Fig. \ref{fig:cvx}. We can treat the arrived message packets as boxes of width $\tau$, and thus the message packets will not interfere if the corresponding boxes do not overlap.  Therefore, we can construct an optimization problem as
\begin{eqnarray}
  \label{eq:jopt}
  \begin{aligned}
    & \underset{\{\Delta_i\}}{\text{minimize }}\max_{i,k}
    & & \Delta_i + \delta_{ki} ,\\
    & \text{subject to}
    & & |\Delta_i + \delta_{ki} - \Delta_{j} + \delta_{kj}| \ge \tau, \\
    & & & \Delta_{i}\ge 0,  \\
  \end{aligned}\\
\forall\mb{ } i,j,k \in [1,2,\ldots,N] \suchthat i,j \ne k \mb{ and } i\ne j. \nonumber 
\end{eqnarray}
This formulation is similar to the optimization problem discussed in \cite{Park-AllToAll}. In  \cite{Park-AllToAll}, a sub-optimal solution for \eqref{eq:jopt} using a heuristic algorithm is proposed, in which the first constraint is enforced, such that instead of all $j\ne i$, only $j=1, 2, \ldots, i-1$ are considered to solve for $\Delta_i$, one by one. Starting with $\Delta_1=0$,  and for each $i$ from $2$ to $N$, smallest $\Delta_i\ge 0$ needed to avoid collision of packets from Node-$j$ (whose delays are already known from previous iteration as $j=1, \ldots, i-1$) is computed. In contrast to \cite{Park-AllToAll}, our proposed algorithms solves the above optimization problem using convex method, TSP and iterative path-adjusting methods.

\subsection{Convex algorithm (CA)}
\label{subsec:CVX}
\begin{figure}[t]
  \centering
  \includegraphics[width=2.5in]{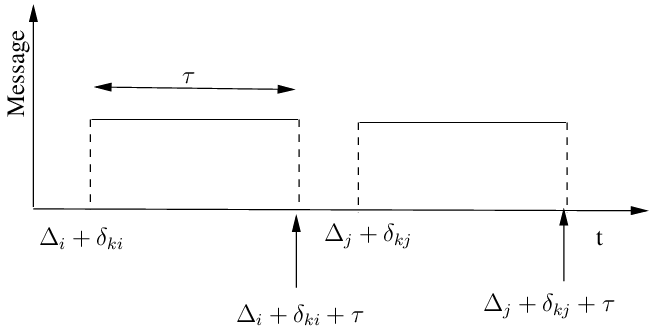}
  \caption{Messages from node $i$ and $j$ arriving at node $k$}
  \label{fig:cvx}
\end{figure}

The optimization problem defined in \eqref{eq:jopt} is not a convex problem since the equality constraints are not affine. The problem can however be made convex by introducing additional constraints. If we have predetermined the order in which the message packets should arrive at a particular node, we can make sure that the corresponding boxes do not overlap using a simple linear inequality. For example, in Fig.~\ref{fig:cvx}, the inequality would be $\Delta_i+\delta_{ki}+\tau\le\Delta_j+\delta_{kj}$. Thus, we have isolated the non-convexity of the optimization problem into selecting the order in which the message packets should arrive at the different nodes. Suppose, we consider a sequential schedule, in which node $i+1$, will transmit after node $i$, then we can construct the optimization problem as

\begin{align}
& \underset{\{\Delta_i\}}{\text{minimize }}\max_{i,k}
& & \Delta_i + \delta_{ki} \label{eq:convex}\\
& \text{subject to}
& & \Delta_i + \delta_{ki} + \tau \le \Delta_{i+1} + \delta_{k,i+1} \label{eq:convex:constraint}\\
& & & \Delta_{i}\ge 0 \label{eq:convex:nonnegative}
\end{align}
In \eqref{eq:convex:constraint}, $i$ goes from 1 to $N-1$, as there is no node with index $N+1$, and $k \neq i,i+1$.

This is a convex optimization problem, as the objective function is convex, and all the inequality constraints are convex. The problem can be solved as a general linear program \cite{Bertsimas-linearopt, Boyd-Book}, but algorithms with lower complexity can be constructed by exploiting the structure of the problem. We found that the most efficient way to solve \eqref{eq:convex} is to minimize the delays $\Delta_i$ sequentially in order of increasing $i$. We note that $\Delta_{i+1}$ is minimized when it is zero or when \eqref{eq:convex:constraint} is tight for at least one $k$. For the first node, the smallest possible delay is $\Delta_1 = 0$ and for subsequent nodes the smallest possible delays are given by
\begin{equation}
\label{eq:convex-delays}
\Delta_{i+1}  = \max\left\{0, \Delta_i + \max_k\left\{\delta_{ki}-\delta_{k,i+1}\right\} + \tau\right\}.
\end{equation}
This results in a solution where none of the delays can be decreased without violating either \eqref{eq:convex:constraint} or \eqref{eq:convex:nonnegative}, meaning that we have found an optimum of \eqref{eq:convex}. The algorithm can be thought of as sliding the boxes corresponding to transmission $i+1$ to the left along the time axis until one of them hits 0 or a box from transmission $i$.

In the above formulation, we have only considered interference between messages from nodes which come directly after each other in the node order. Given that node $i$ does not receive a message from itself, it may be possible for messages from node $i-1$ and node $i+1$ to interfere when they are received at node $i$. This can however never happen, as \eqref{eq:convex:constraint} implies that
\begin{equation}
\label{eq:nocollision}
\Delta_{i-1} + \delta_{i,i-1} + \tau \leq \Delta_{i+1} + \delta_{i,i+1}
\end{equation}
for $i = 2,3,\ldots,N-1$. This is shown in Appendix A. Given that $N$ delays need to be computed and that $N$ path delays must be considered in each computation, the algorithm has a complexity of $O(N^2)$.

If the node order is set to A, B, C, in the configuration defined in Table~\ref{tab:cfg}, this method produces the same solution as the grid search method, within the grid search tolerance\footnote{Note that the convex solution does not always produce the optimal solution, and thus may not always match the result from the grid search method.}. Even though the formulated problem is convex, for the $N$-node scenario, sequential ordering may not be the optimal order with the lowest report cycle. Selecting an optimal order is in itself  a combinatorial optimization problem \cite{Lawler-combopt,lovasz-comb}. However, for most practical scenarios, we can select an arbitrary order and solve the convex problem as demonstrated in Section \ref{sec:SS}.

If we consider a configuration of \cite{Ramezani} for the B-MAC broadcast packet scheduling problem, with only anchors in a fully-connected single-channel configuration, then the proposed B-MAC scheduling problem is similar to the optimization problem of section IV-A. However, they are not exactly the same. For example, the constraint (7) in  \cite{Ramezani}, is more stringent as it ensures that the the acoustic signals in the UWA network does not collide in the space between anchors nodes. This is not applicable for electromagnetic wave based WSN networks as there is no separation between the anchor-node and ordinary sensor-node in the system and only interference at all the node locations need to be nulled. This is accomplished efficiently through the constraint defined in (12).  \newline

\begin{figure*}[!t]
  \centering
  \subfloat[Cyclic order $\ldots-B-A-C-B-\ldots$]{
  \begin{tikzpicture}

    \colorlet{color1}{green}
    \colorlet{color2}{red}
    \colorlet{color3}{blue}

    \tikzstyle{pulse} = [ultra thick]
    \tikzstyle{graypulse} = [ultra thick, densely dashed, opacity=0.5]
    \tikzstyle{transmission} = [opacity=1]
    \tikzstyle{graytransmission} = [dashed, opacity=0.5]

    \begin{scope}[x={(\xunit,0pt)}, y={(0pt,\yunit)}]

      \draw[black,->] (-50,0) -- (1050,0);
      \foreach \t in {0,100,...,1000}
         \draw (\t,0) -- (\t,-0.075) node[font=\footnotesize, anchor=north, inner sep=0pt, shift={(0,-0.05)}] {$\t$};
      \node[font=\footnotesize, anchor=north, inner sep=0pt] at (500,-0.4) {Time [ns]};

      \draw (-50,0) -- (-50,4);
      \node[font=\footnotesize, anchor=south, rotate=90, color1, inner sep=0pt, yshift=3] at (-50,3) {Node A};
      \node[font=\footnotesize, anchor=south, rotate=90, color2, inner sep=0pt, yshift=3] at (-50,2) {Node B};
      \node[font=\footnotesize, anchor=south, rotate=90, color3, inner sep=0pt, yshift=3] at (-50,1) {Node C};

      \draw[graypulse, color3] (-50.000,3) -- (16.667,3);
      \draw[graypulse, color3] (-50.000,2) -- (33.333,2);
      \draw[graypulse, color1] (-50.000,1) -- (-33.333,1);
      \draw[graypulse, color2] (16.667,3) -- (116.667,3);
      \draw[graypulse, color3] (216.667,3) -- (316.667,3);
      \draw[graypulse, color1] (133.333,2) -- (233.333,2);
      \draw[graypulse, color3] (233.333,2) -- (333.333,2);
      \draw[graypulse, color1] (166.667,1) -- (266.667,1);
      \draw[graypulse, color2] (66.667,1) -- (166.667,1);
      \draw[pulse, color2] (316.667,3) -- (416.667,3);
      \draw[pulse, color3] (516.667,3) -- (616.667,3);
      \draw[pulse, color1] (433.333,2) -- (533.333,2);
      \draw[pulse, color3] (533.333,2) -- (633.333,2);
      \draw[pulse, color1] (466.667,1) -- (566.667,1);
      \draw[pulse, color2] (366.667,1) -- (466.667,1);
      \draw[graypulse, color2] (616.667,3) -- (716.667,3);
      \draw[graypulse, color3] (816.667,3) -- (916.667,3);
      \draw[graypulse, color1] (733.333,2) -- (833.333,2);
      \draw[graypulse, color3] (833.333,2) -- (933.333,2);
      \draw[graypulse, color1] (766.667,1) -- (866.667,1);
      \draw[graypulse, color2] (666.667,1) -- (766.667,1);
      \draw[graypulse, color2] (916.667,3) -- (1016.667,3);
      \draw[graypulse, color1] (1033.333,2) -- (1050.000,2);
      \draw[graypulse, color2] (966.667,1) -- (1050.000,1);

      \draw[transmission, color1] (116.667,0) -- (116.667,3.3);  
      \draw[transmission, color2] (0.000,0) -- (0.000,3.3);  
      \draw[transmission, color3] (166.667,0) -- (166.667,3.3);  
      \draw[graytransmission, color1] (416.667,0) -- (416.667,4);
      \draw[graytransmission, color2] (300.000,0) -- (300.000,3.3);  
      \draw[graytransmission, color3] (466.667,0) -- (466.667,4);
      \draw[graytransmission, color1] (716.667,0) -- (716.667,4);
      \draw[graytransmission, color2] (600.000,0) -- (600.000,4);
      \draw[graytransmission, color3] (766.667,0) -- (766.667,4);
      \draw[graytransmission, color1] (1016.667,0) -- (1016.667,4);
      \draw[graytransmission, color2] (900.000,0) -- (900.000,4);

      \node[color1, fill=white, font=\tiny, inner sep=0pt, anchor=center] at (116.667,0.5) {$\Delta_A$};
      \node[color2, fill=white, font=\tiny, inner sep=0pt, anchor=center] at (0.000,0.5) {$\Delta_B$};
      \node[color3, fill=white, font=\tiny, inner sep=0pt, anchor=center] at (166.667,0.5) {$\Delta_C$};

      \draw[decorate, decoration=brace, font=\footnotesize] (0.000, 3.3) --  (116.667, 3.3) node[midway, anchor=south, yshift=1pt] {$c_{\mathsmaller{BA}}$} ;
      \draw[decorate, decoration=brace, font=\footnotesize] (116.667, 3.3) --  (166.667, 3.3) node[midway, anchor=south, yshift=1pt] {$c_{\mathsmaller{AC}}$} ;
      \draw[decorate, decoration=brace, font=\footnotesize] (166.667, 3.3) --  (300.000, 3.3) node[midway, anchor=south, yshift=1pt] {$c_{\mathsmaller{CB}}$} ;
      \draw[decorate, decoration=brace, font=\footnotesize] (0.000, 3.65) --  (300.000, 3.65) node[midway, anchor=south, yshift=1pt] {total TSP cost} ;

    \end{scope}
  \end{tikzpicture}
  \label{fig:cyclic}}\hfill
  \subfloat[Linear order $B-A-C$]{
  \begin{tikzpicture}

    \colorlet{color1}{green}
    \colorlet{color2}{red}
    \colorlet{color3}{blue}

    \tikzstyle{pulse} = [ultra thick]
    \tikzstyle{transmission} = [opacity=1]

    \begin{scope}[x={(\xunit,0pt)}, y={(0pt,\yunit)}]

      \draw[black,->] (-50,0) -- (1050,0);
      \foreach \t in {0,100,...,1000}
         \draw (\t,0) -- (\t,-0.075) node[font=\footnotesize, anchor=north, inner sep=0pt, shift={(0,-0.05)}] {$\t$};
      \node[font=\footnotesize, anchor=north, inner sep=0pt] at (500,-0.4) {Time [ns]};

      \draw (-50,0) -- (-50,4);
      \node[font=\footnotesize, anchor=south, rotate=90, color1, inner sep=0pt, yshift=3] at (-50,3) {Node A};
      \node[font=\footnotesize, anchor=south, rotate=90, color2, inner sep=0pt, yshift=3] at (-50,2) {Node B};
      \node[font=\footnotesize, anchor=south, rotate=90, color3, inner sep=0pt, yshift=3] at (-50,1) {Node C};

      \draw[pulse, color2] (316.667,3) -- (416.667,3);
      \draw[pulse, color3] (516.667,3) -- (616.667,3);
      \draw[pulse, color1] (433.333,2) -- (533.333,2);
      \draw[pulse, color3] (533.333,2) -- (633.333,2);
      \draw[pulse, color1] (466.667,1) -- (566.667,1);
      \draw[pulse, color2] (366.667,1) -- (466.667,1);

      \draw[transmission, color1] (116.667,0) -- (116.667,4);
      \draw[transmission, color2] (0.000,0) -- (0.000,4);
      \draw[transmission, color3] (166.667,0) -- (166.667,4);
      
      \node[color1, fill=white, font=\tiny, inner sep=0pt, anchor=center] at (116.667,0.5) {$\Delta_A$};
      \node[color2, fill=white, font=\tiny, inner sep=0pt, anchor=center] at (0.000,0.5) {$\Delta_B$};
      \node[color3, fill=white, font=\tiny, inner sep=0pt, anchor=center] at (166.667,0.5) {$\Delta_C$};

    \end{scope}
  \end{tikzpicture}
  \label{fig:bac}} \\
  \subfloat[Linear order $A-C-B$]{
  \begin{tikzpicture}

    \colorlet{color1}{green}
    \colorlet{color2}{red}
    \colorlet{color3}{blue}

    \tikzstyle{pulse} = [ultra thick]
    \tikzstyle{transmission} = [opacity=1]

    \begin{scope}[x={(\xunit,0pt)}, y={(0pt,\yunit)}]

      \draw[black,->] (-50,0) -- (1050,0);
      \foreach \t in {0,100,...,1000}
         \draw (\t,0) -- (\t,-0.075) node[font=\footnotesize, anchor=north, inner sep=0pt, shift={(0,-0.05)}] {$\t$};
      \node[font=\footnotesize, anchor=north, inner sep=0pt] at (500,-0.4) {Time [ns]};

      \draw (-50,0) -- (-50,4);
      \node[font=\footnotesize, anchor=south, rotate=90, color1, inner sep=0pt, yshift=3] at (-50,3) {Node A};
      \node[font=\footnotesize, anchor=south, rotate=90, color2, inner sep=0pt, yshift=3] at (-50,2) {Node B};
      \node[font=\footnotesize, anchor=south, rotate=90, color3, inner sep=0pt, yshift=3] at (-50,1) {Node C};

      \draw[pulse, color2] (500.000,3) -- (600.000,3);
      \draw[pulse, color3] (400.000,3) -- (500.000,3);
      \draw[pulse, color1] (316.667,2) -- (416.667,2);
      \draw[pulse, color3] (416.667,2) -- (516.667,2);
      \draw[pulse, color1] (350.000,1) -- (450.000,1);
      \draw[pulse, color2] (550.000,1) -- (650.000,1);

      \draw[transmission, color1] (0.000,0) -- (0.000,4);
      \draw[transmission, color2] (183.333,0) -- (183.333,4);
      \draw[transmission, color3] (50.000,0) -- (50.000,4);
      
      \node[color1, fill=white, font=\tiny, inner sep=0pt, anchor=center] at (0.000,0.5) {$\Delta_A$};
      \node[color2, fill=white, font=\tiny, inner sep=0pt, anchor=center] at (183.333,0.5) {$\Delta_B$};
      \node[color3, fill=white, font=\tiny, inner sep=0pt, anchor=center] at (50.000,0.5) {$\Delta_C$};

    \end{scope}
  \end{tikzpicture}
  \label{fig:acb}}\hfill
  \subfloat[Linear order $C-B-A$]{
  \begin{tikzpicture}

    \colorlet{color1}{green}
    \colorlet{color2}{red}
    \colorlet{color3}{blue}

    \tikzstyle{pulse} = [ultra thick]
    \tikzstyle{transmission} = [opacity=1]

    \begin{scope}[x={(\xunit,0pt)}, y={(0pt,\yunit)}]

      \draw[black,->] (-50,0) -- (1050,0);
      \foreach \t in {0,100,...,1000}
         \draw (\t,0) -- (\t,-0.075) node[font=\footnotesize, anchor=north, inner sep=0pt, shift={(0,-0.05)}] {$\t$};
      \node[font=\footnotesize, anchor=north, inner sep=0pt] at (500,-0.4) {Time [ns]};

      \draw (-50,0) -- (-50,4);
      \node[font=\footnotesize, anchor=south, rotate=90, color1, inner sep=0pt, yshift=3] at (-50,3) {Node A};
      \node[font=\footnotesize, anchor=south, rotate=90, color2, inner sep=0pt, yshift=3] at (-50,2) {Node B};
      \node[font=\footnotesize, anchor=south, rotate=90, color3, inner sep=0pt, yshift=3] at (-50,1) {Node C};

      \draw[pulse, color2] (450.000,3) -- (550.000,3);
      \draw[pulse, color3] (350.000,3) -- (450.000,3);
      \draw[pulse, color1] (566.667,2) -- (666.667,2);
      \draw[pulse, color3] (366.667,2) -- (466.667,2);
      \draw[pulse, color1] (600.000,1) -- (700.000,1);
      \draw[pulse, color2] (500.000,1) -- (600.000,1);

      \draw[transmission, color1] (250.000,0) -- (250.000,4);
      \draw[transmission, color2] (133.333,0) -- (133.333,4);
      \draw[transmission, color3] (0.000,0) -- (0.000,4);
      
      \node[color1, fill=white, font=\tiny, inner sep=0pt, anchor=center] at (250.000,0.5) {$\Delta_A$};
      \node[color2, fill=white, font=\tiny, inner sep=0pt, anchor=center] at (133.333,0.5) {$\Delta_B$};
      \node[color3, fill=white, font=\tiny, inner sep=0pt, anchor=center] at (0.000,0.5) {$\Delta_C$};

    \end{scope}
  \end{tikzpicture}
  \label{fig:cba}}
  \caption{The cyclic TSP solution \protect\subref{fig:cyclic} and the 3 possible linear orders that can be created from it \protect\subref{fig:bac}-\protect\subref{fig:cba}, for the network in Fig. \ref{fig:p2p}. The filled in horizontal bars show one cycle of received messages. The times of transmission are shown as solid vertical lines. In previous and future cycles, received messages and times of transmission are shown as dashed bars and dashed lines respectively. The TSP-costs along the cheapest tour can be visualised as the time differences between the transmissions. The duration of one cycle is 30 ns. In \protect\subref{fig:bac}, \protect\subref{fig:acb}, and \protect\subref{fig:cba}, the report cycles are approximately 633 ns, 650 ns, and 700 ns respectively.}
\label{fig:tsp}
\end{figure*}
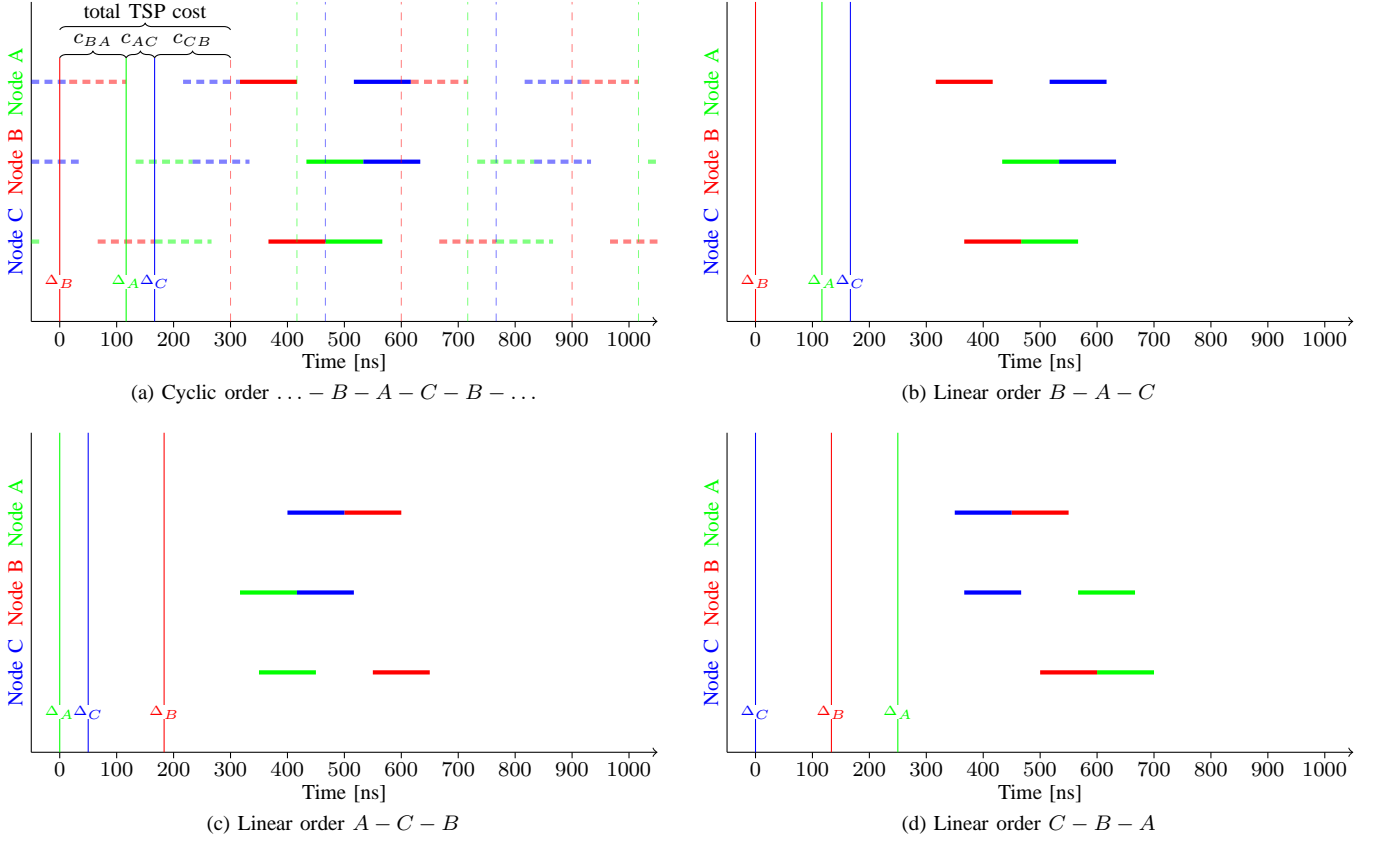

\subsection{Optimizing the node order by solving a TSP}
\label{subsec:tsp}
The problem of selecting a good node order can be formulated as an asymmetric traveling salesman problem (TSP) \cite{Lawler-tour}, where the cost matrix is derived from the path delays. To be able to do this, we modify the problem so that the nodes transmit in a cyclic order where a second message from the first node is placed directly after the first message from the last node. Then we solve a TSP problem which minimizes the time between two transmissions from the same node. In \cite{Chen-CSMA,LMAC} authors discuss the broadcast schedule construction for a UWA network using the traveling salesman approach, here anchor-nodes cannot transmit before they have received the message from the previous node in the sequence to avoid collision of acoustic waves at sensor nodes. This will result in a symmetric TSP problem. In the context of a general all-to-all communication in WSN this constraint is not need and this will manifest the problem as an asymmetric TSP problem as shown later. Finally, we consider the $N$ different ways in which the cyclic order can be broken into a linear order, and select the alternative which minimizes the report cycle in the original problem.

The objective of the traveling salesman problem is to find the cheapest tour which visits a number of cities exactly once. The input to the problem is a cost matrix, $\mbf{C}$, where its element $c_{ij}$ is the cost of going from city $i$ to city $j$ \cite{Lawler-tour}. In our problem, we let each city correspond to a node in the network.  We define the cost matrix so that $c_{ij}$ is the minimum difference between the delays of node $j$ and node $i$, allowed by \eqref{eq:convex:constraint}, given that $j$ comes directly after $i$ in the node order. Given that we are looking at a cyclic order, node $1$ takes the role of node $N+1$ in \eqref{eq:convex:constraint}, and we do not need to take the constraints \eqref{eq:convex:nonnegative} into consideration. If node $j$ comes directly after node $i$ in the selected order, we have that
\begin{equation}
\Delta_{j} = \Delta_i + \max_k\left\{\delta_{ki}-\delta_{kj}\right\} + \tau.
\end{equation}
In other words, the delay of any node is equal to the delay of the previous node, plus the cost
\begin{equation}
\label{eq:tsp-costs}
c_{ij} = \max_k\left\{\delta_{ki}-\delta_{kj}\right\} + \tau.
\end{equation}
By adding up all of the costs associated with the successive node pairs in the transmission order (TSP tour), we therefore get the time between two transmissions made by the same node. The problem of minimizing the time between two transmissions made by the same node can therefore be formulated as a TSP where the cost matrix is defined by $c_{ij}$. For the $3$ node configuration shown in Fig.~\ref{fig:p2p}, the algorithm is graphically illustrated in Fig.~\ref{fig:tsp}. 

The TSP is known to be NP-hard \cite{Ramaswamy-NP,Lawler-combopt,lovasz-comb}, but there are algorithms that can find exact solutions for small problems, and other algorithms that can find approximate solutions for larger problems \cite{Johnson-np,Laporte-tspsoln}. Many techniques employ heuristic approaches for finding the approximate solution \cite{helsgaun2000effective,Reeves:modernHeu,Rosenkrantz-heu}. The best approximate algorithms, often produce optimal or very close to optimal solutions, for large networks with hundreds of nodes. Furthermore, the approximate algorithms can be run multiple times with different starting points and thereby achieve much better performance \cite{Perttunen-InitHeu}. We have chosen to use the TSP solver LKH \cite{helsgaun2000effective}, which is based on the Lin-Kernighan heuristic.  For a problem with 100 nodes, LKH requires less than a second to produce a solution which has a high probability of being optimal. In LKH, all of the costs in matrix $\mbf{C}$, must be integers and therefore we mapped the costs in each problem to the the interval between 0 and $10^6$ using an affine mapping and rounded them to the closest integers. We used version 2.0.7 of LKH with the default settings for all problems.

The report cycle will depend on which node in the TSP cycle is selected as node $1$. Therefore we consider all of the $N$ possible choices for node $1$ and solve the convex problem defined in Section \ref{subsec:CVX} for each one of them to see which alternative results in the shortest report cycle. Given that we can reuse the costs that we computed in \eqref{eq:tsp-costs} when we compute the delays in \eqref{eq:convex-delays}, the problem of choosing a first node has complexity $O(N^2)$. The overall complexity is therefore dominated by the TSP solver, which has an average complexity that scales approximately as $O(N^{2.2})$ \cite{helsgaun2000effective}. We may introduce some sub-optimality by transforming the problem into a problem with transmitters in a cyclic order, and LKH may also not find the exact optimum of the TSP. The gap to optimality would however be negligible for most practical applications.


\subsection{Iterative path-adjusting algorithm (IPA)}
\label{subsec:IA}
The CA reduces the algorithm complexity by allowing sub-optimality due to the fixed ordering. On the other hand, the TSP algorithm improves over the CA, by choosing a better order without increasing the algorithmic complexity on average. One problem with both the algorithms is their inefficiency when \eqref{eq:con} holds for most of the nodes (i.e., nodes are scattered far-apart compared to $\mathcal{L}$). For a random node configuration, we can in theory ensure that \eqref{eq:con} holds by making the message length $\tau$ small enough. If $\tau$ is decreased by $d\tau$, the report cycle of the algorithms will however only decrease by $Nd\tau$, as the algorithms cannot change the order in which messages are to be received at the nodes. This results in poor performance when $\tau$ is small in comparison to the path delays of the network. To overcome this problem, we propose an alternative algorithm called iterative path-adjusting algorithm (IPA). We show in the later sections that this algorithm outperforms the convex formulation with strict ordering as defined in \eqref{eq:convex:constraint}, and the TSP algorithm, when the sensor nodes are scattered wide apart.

In this algorithm, we adjust the path differences between nodes, $d_{ki}$ and $d_{kj}$ to satisfy \eqref{eq:con} in an iterative way. Adjusting the path difference is the same as introducing delays at nodes $i$ and $j$, so that the signals from $i$ and $j$ do not interfere at node $k$. The algorithm is described below in three steps followed by an example on a 3-node network.

\begin{figure}[!t]
  \centering
  \includegraphics[width=20pc]{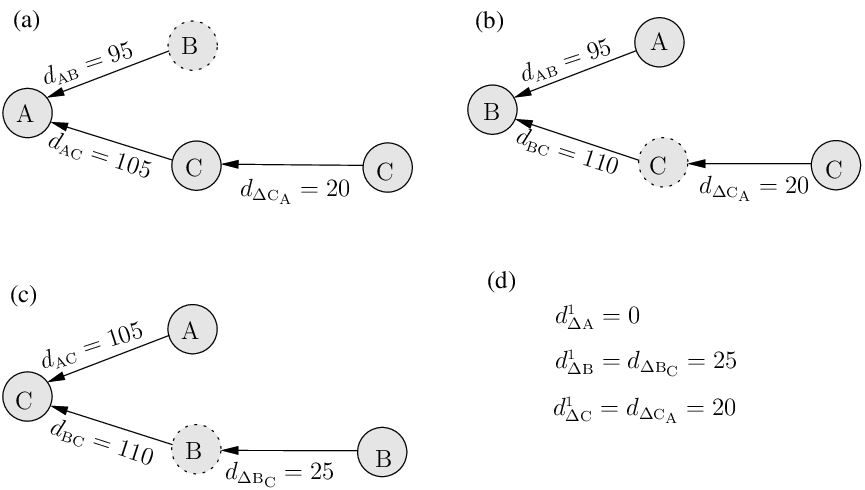}
  \caption{Iteration 1 for  the $3$ node network shown in Fig. \ref{fig:p2p} with the configuration defined in Table \ref{tab:cfg}.}
  \label{fig:iter1}
\end{figure}

\begin{itemize}
\item[1] Start the first iteration with $l=0$ ($l+1$ denotes the iteration number) and $k=1$, with $d_{ki}^0=d_{ki}$. For a topology having $N$ nodes, add additional path lengths $d_{\Delta i_k}^{l+1}$ and $d_{\Delta j_k}^{l+1}$, $\forall i,j \in [1,2,\ldots,N],\mb{ } i,j \ne k \mb{, } i \ne j$  to nodes  $i$ and $j$ to satisfy \eqref{eq:con}. Thus, the new path lengths are given by
\begin{eqnarray}
  \label{eq:pathadj}
   d_{ki}^{l+1}&=&d_{ki}^l+d_{\Delta i_k}^{l+1},\\
   d_{kj}^{l+1}&=&d_{kj}^l+d_{\Delta j_k}^{l+1}.
\end{eqnarray}
Note that to satisfy \eqref{eq:con}, the path-length needs to be added to one of the nodes $i$ or $j$. In this algorithm, we set  $d_{\Delta   i_k}^{l+1}=0$ and add additional path length $d_{\Delta j_k}^{l+1}$  only to node $j$.
\item[2] Repeat Step 1, by selecting all nodes one by one ($k=1,2,\ldots,N$) in the network. Each time, carry over additional path lengths added  $d_{ki}^l+d_{\Delta i_k}^{l+1}$ and $d_{kj}^l+d_{\Delta j_k}^{l+1}$. The total adjusted path lengths at the end of iteration $l$ are given by
\begin{equation}
  \label{eq:sumpath}
  d_{\Delta k}^{^{l+1}}=d_{\Delta k}^l+\sum_i d_{\Delta i_k}^{l+1},
\end{equation}
for $k\in [1,2,\ldots,N]$ and $k\ne i$. This completes an iteration.
\item[3] Repeat Step 1 and Step 2 until the total adjusted path length for each node does not change across iterations, meaning that the following condition holds for all $k \in [1,2,\ldots,N]$.
\begin{equation}
  \label{eq:d}
  d_{\Delta k}^{l+1}=d_{\Delta k}^l.
\end{equation}

This indicates that \eqref{eq:con} is met for all nodes simultaneously.\\
\end{itemize}
The proposed method  is illustrated with the $3$ node network shown in Fig. \ref{fig:p2p}, with the configuration defined in Table~\ref{tab:cfg}.
Fig.~\ref{fig:iter1} (a) shows that the addition of an additional path length of $20=(30-(d_{\MS{AC}}-d_{\MS{AB}}))$ is required at node $C$ to meet the constraints in \eqref{eq:con} so that signals from $B$ and $C$ do not collide at $A$. Similarly, Fig.  \ref{fig:iter1} (b) and  Fig. \ref{fig:iter1} (c) add additional path lengths to the previous topology to avoid collisions at nodes $B$ and $C$ respectively. At the end of the 1st iteration, the total path lengths for all of the nodes are given in Fig.  \ref{fig:iter1} (d).

\begin{figure}[!t]
  \centering
  \includegraphics[width=20pc]{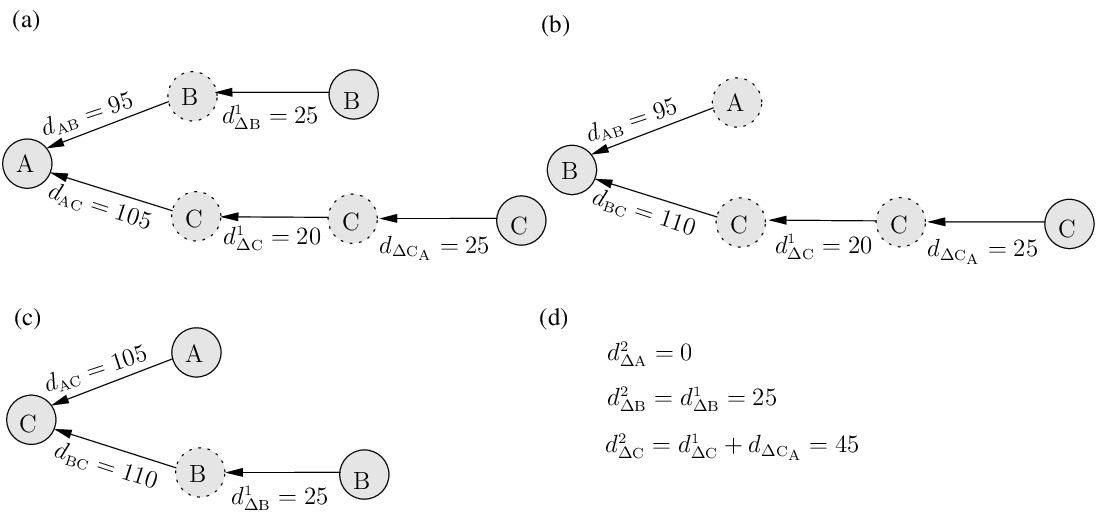}
  \caption{Iteration 2 for  the $3$ node network shown in Fig. \ref{fig:p2p} with the configuration defined in Table \ref{tab:cfg}.}
  \label{fig:iter2}
\end{figure}

The second iteration is illustrated in Fig. \ref{fig:iter2}. Notice that we carried the new topology with added path lengths  from the previous iteration $(d_{\Delta A}^1, d_{\Delta B}^1,  d_{\Delta C}^1)$ to iteration $2$ and at the end of iteration $2$, the total added path lengths are ($d_{\Delta A}^2, d_{\Delta B}^2, d_{\Delta C}^2)=(0,25,45)$. Now the iteration is stopped as it meets the conditions defined in Step 3.

Translating the path lengths into path delays by dividing by the speed of light, $c$, results in $(\Delta_{A},\Delta_{B},\Delta_{C})\approx(0\,,84\,,150)~\mb{[ns]}$. This is the same result as with the grid search and convex methods. This  algorithm is analyzed in the next Section.

\subsection{Analysis of IPA}
\label{sec:CCIA}
In order for the arriving signals not to interfere, \eqref{eq:con} needs to be satisfied. We define the path matrix, $\mbf{M}$, where each element of $\mbf{M}$, $d_{ki}$, denotes the distance between node $k$ and node $i$. The IPA adjusts the path matrix in such a way that  the path lengths to node $k$ from other nodes (represented by the $k$-th row in the matrix $\mbf{M}$), have path differences greater than $\mathcal{L}$. For an $N$ node network, the algorithm  starts with the original path matrix, $\mbf{M}^0$, as given in \eqref{eq:app2}. The path adjusted matrix after the $l$-th iteration is represented as $\mbf{M}^l$.
\begin{equation}
\label{eq:app2}
\mbf{M}^0= \begin{bmatrix}
  0 & d_{12} & \cdots & d_{1N} \\
  d_{21} & 0 & \cdots & d_{2N} \\
  \vdots  & \vdots  & \ddots & \vdots  \\
  d_{N1} & d_{N2} & \cdots & 0
 \end{bmatrix}.
\end{equation}
Note that $d_{kk}=0 \mbox{, }\forall k\in (1,2,\ldots,N)$. Also, matrix $\mbf{M}^0$ is symmetric, that is $d_{ki}=d_{ik}$.  The path adjusted matrix, $\mbf{M}^l$, has elements, $d_{ij}^l$. The $i$-th row of the path adjusted matrix $\mbf{M}^l$ is denoted as
\begin{equation}
  \underline{d_{ix}^l}=
  \begin{bmatrix}
  d_{i1}^l & d_{i2}^l & \cdots & d_{iN}^l
  \end{bmatrix},
\end{equation}
similarly, the $i$-th column is denoted as
\begin{equation}
  \underline{d_{xi}^l}=
  \begin{bmatrix}
  d_{1x}^l & d_{2x}^l & \cdots & d_{Nx}^l
  \end{bmatrix}^{\mbox{\scriptsize T}},
\end{equation}
where $\mbox{\scriptsize T}$ denotes the transpose operator.

To perform step 1 of the algorithm, there are many possibilities for  additional path lengths $d_{\Delta i_k}^l$ and $d_{\Delta j_k}^l$, such that the arriving signals  at node $k$, have effective path length difference greater than $\mathcal{L}$. In this paper, in order to make the arriving signals to node $k$ from nodes $i$ and $j$ satisfy $|d_{ki}^l-d_{kj}^l|\ge \mathcal{L}$, we will add path lengths only to $j$, if $j>i$. That is,
\begin{flalign}
  \label{eq:if}
  &\mb{if }|d_{ki}^l-d_{kj}^l|<\mathcal{L}\mb{ and }j>i \mb{ then}&
\end{flalign}
\begin{align}
&d_{\Delta i_k}^{l+1}=0, \label{eq:app3}\\
&d_{\Delta j_k}^{l+1}=\mathcal{L}-(d_{kj}^l-d_{ki}^l),\label{eq:app3a}\\
&\underline{d_{xj}^{l+1}}=[\underline{d_{xj}^{l}}+d_{\Delta j_k}^{l+1}\underline{1}] ,\label{eq:app4} \\
&\forall i,j \in [1,2,\ldots,N],\mb{ } i,j \ne k,  i \ne j,  \mb{ and } j > i.\nonumber
\end{align}
Where, \underline{1} is $[1,1,\ldots,1]^{\mb{\scriptsize T}}$  and the  process, defined in  \eqref{eq:if} to \eqref{eq:app4} is repeated for $k=1,2,\ldots,N $ sequentially to complete an iteration. The iterations with $l=0,1,2,\ldots $ are performed until in \eqref{eq:app3a}, $d_{\Delta j_k}^{l+1}=0, \forall i,j,k\in [1,2,\ldots,N],\mb{ } i,j \ne k \mb{ and }  i \ne j$ is met.  At each iteration, the elements from $\mbf{M}^{l}$  are used for \eqref{eq:if} to \eqref{eq:app4}.

During each iteration, when \eqref{eq:if} is met, the additional path is added only to one of the nodes (the node on the right). Therefore, as the iterations increase, the path adjusted matrix, $\mbf{M}$,  will converge to the state with its elements
\begin{equation}
  \label{eq:app9}
\begin{split}
  &|d_{ki}^{l^*}-d_{kj}^{l^*}|\ge\mathcal{L}, \\
 &\forall i,j,k \in [1,2,\ldots,N], i,j\ne k \mb{ and } i\ne j,
\end{split}
\end{equation}
where, $l^*+1$, denotes the number of iterations required for convergence. At this state the arriving signals to any node $k$ from nodes $i$ and $j$ will satisfy \eqref{eq:con}. The effective adjusted path is given by
\begin{equation}
  \label{eq:app10}
  d_{\Delta i}^{l^*}=\underline{d_{1x}^{l^*}}(i)-\underline{d_{1x}^0}(i),
\end{equation}
and the equivalent added delay for node, $i$, is $\Delta_i=d_{\Delta i}^{l^*}/\mu$.


The average algorithmic complexity for IPA is evaluated by a least square polynomial fit to the average computational time (in ticks) consumed by the algorithm for networks of different sizes. The procedure followed, along with the results, are discussed  in Appendix B. From the results of the Appendix B, the average complexity of the IPA algorithm is $O(N^3)$.
 
A summary of the average complexities of the proposed methods is shown in Table \ref{tab:summary}. For CA, the average case and the worst case complexity are the same, therefore in networks, where the real-time guaranties are needed CA is more amenable than TSP and IPA. The IPA opens up for a higher flexibility regarding the order of the transmissions. Thus, it provides better throughput  compared to the convex algorithm as shown in Section \ref{sec:SS}. In mobile sensor networks, the convex approach with fixed ordering among the nodes opens up for schedule-based communication and ranging; thus, node information need not be encoded in the packets \cite{Satyam-Sch-letter}. On the other hand, IPA requires transmission overhead since the node information has to be included in the packets, as the order of packet reception is not predetermined. However, the overhead of encoding the node information in the packet ($\log_2N$ bits) is not significant.

The performances of the orthrogonalization, CA, TSP, and IPA under large scale networks with different geometric  formations are studied in the Section \ref{sec:SS}.

\begin{table}[!t]
  \caption{Summary of average complexity of the proposed methods.}
  \centering
  \begin{tabular}{|l|c|}
    \hline
    \textbf{Algorithm} & \textbf{Average case complexity} \\
    \hline
    \multirow{2}{*} {Grid Search} & $O(q^N)$ \\ & ($q$ is the size of the quantized grid) \\ \hline
    \multirow{2}{*} {CA} & $O(N^2)$ \\ & (After exploiting the structure in the LP problem)\\ \hline
    \multirow{2}{*}{TSP } &  $O(N^{2.2})$ \\ & (Using LKH solver) \\
    \hline
    IPA & $O(N^3)$ \\
    \hline
  \end{tabular}
  \label{tab:summary}
\end{table}

\begin{figure}
  \centering
  \includegraphics[scale=0.5]{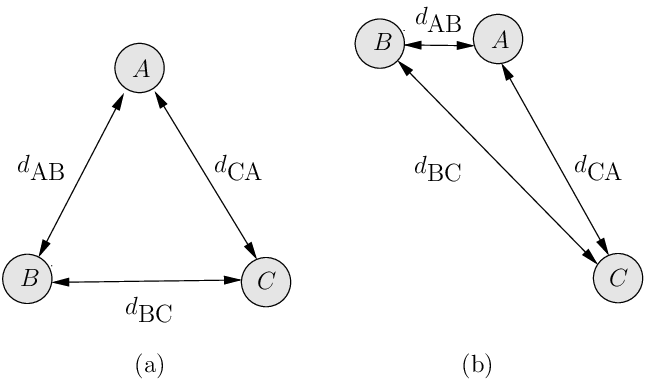}
  \caption{Two distinct $3$-node configurations.}
  \label{fig:PowerCDMA}
\end{figure}

\section{Effect of Syncronization and Range errors}
\label{sec:sync}
In the discussion so far, we assumed that the sensor clocks are synchronized and the available set of range estimates ($d_{ij}$s) are accurate. Accurate network synchronization can be achieved by synchronizing the sensor clocks to  the global positioning system (GPS) clocks or in GPS deficient systems by using the protocols discussed in Section \ref{sec:sys-model}. Accurate range estimation can be accomplished by using TOA methods. The best performance in terms of mean-square-error (MSE) for an unbiased estimator is given by the Cramer Rao lower bound (CRLB) and for a time of arrival (TOA) estimation problem this is given by \cite{gezici-mag,dardari}:
\begin{equation}
  \label{eq:crb}
  \sigma_{\tau}^2\ge \frac{1}{8\pi^2\mb{SNR}\beta^2},
\end{equation}
where $\beta$, is the effective signal bandwidth defined by
\begin{equation}
  \label{eq:beta}
  \beta^2=\left[ \frac{\int_{-\infty}^{\infty}f^2|S(f)|^2df}{\int_{-\infty}^{\infty}|S(f)|^2df}   \right],
\end{equation}
where $S(f)$, is the Fourier transform of the transmit pulse, $s(t)$.
Since many technologies like UWB use extremely large bandwidths, they can be used for precise range estimation.  Practical UWB hardware with ranging and communication capabilities with range estimation accuracy of a few centimeters are discussed in \cite{Ales-2,VJ-3}.

In practice, clock synchronization is not perfect and there will be range errors. These will result in message packets colliding at the receiving nodes. The synchronization error can be approximated as a zero mean normal distribution as shown in \cite{Elson-clocksync,Santashil-clocksync}. In wideband RF systems \cite{Venkat-radar, Hightower-RFreport}, problems such as multi-path fading, background interference, and irregular signal propagation characteristics make range estimates inaccurate. The range error can also be approximated as a zero mean normal distribution \cite{Koen-rangeerror,Lovelace-jitter}.

We assume that synchronization and range errors are independent and the net effect will result in the packet arrival time to be randomly shifted from the intended  position. The distribution of this random shift from the true position can be approximated to a Gaussian distribution, $\mathcal{N}(0,\sigma_e^2)$. This shift in time of arrival of the message packets at the receiving node can cause interference due to packet collisions. This problem can be reduced by adding a guard interval, $\epsilon$, to the equations of the CA, TSP,  and IPA. This can be done by expanding the message packet length $\tau$ to $\tau'=\tau+\epsilon$ in  \eqref{eq:convex:constraint} and \eqref{eq:con}. Where $\epsilon$ can be used to trade off between tolerable interference and the report cycle time (update rate). We can show that
\begin{equation}
  \label{eq:ep}
  \epsilon=\sqrt{2}\sigma_e\mb{erfc}^{-1}(2(1-\mathcal{P})),
\end{equation}
where $1-\mathcal{P}$ denotes the percentage during which neighboring packets collide due to the range and synchronization errors. For example, to have $95\%$ collision avoidance between neighboring packets, we need to have $\epsilon=1.65\sigma_e$. The network level performance in presence of range and synchronization errors, using the above method, for proposed algorithms are studied in Section \ref{sec:SS}.
\begin{figure*}[!t]
  \centering
    \subfloat[Network Topology]{  \includegraphics[scale=0.5]{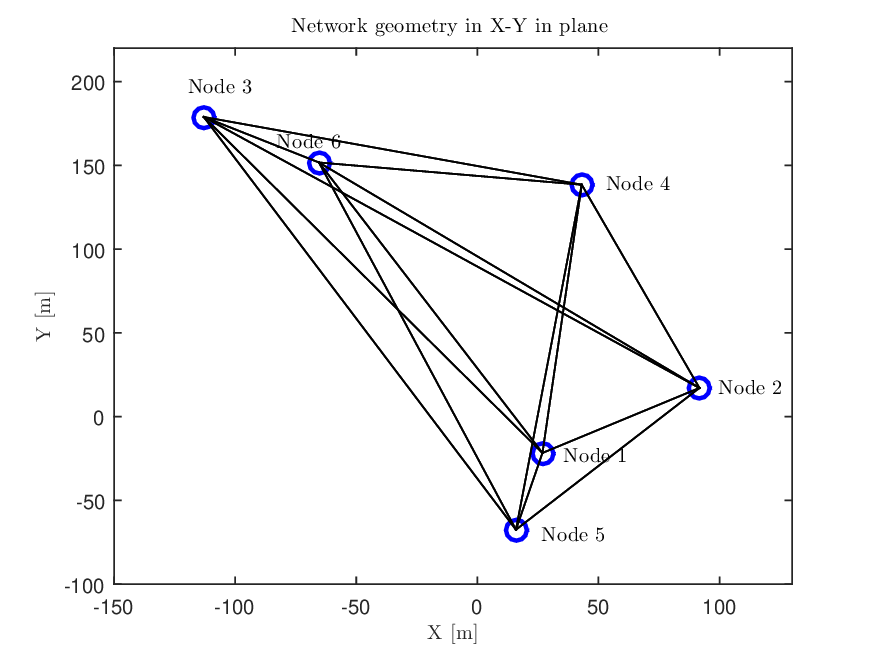} \label{fig:NetworkTopology_without_outliers}}
    \subfloat[Convex Algorithm]{\includegraphics[scale=0.55]{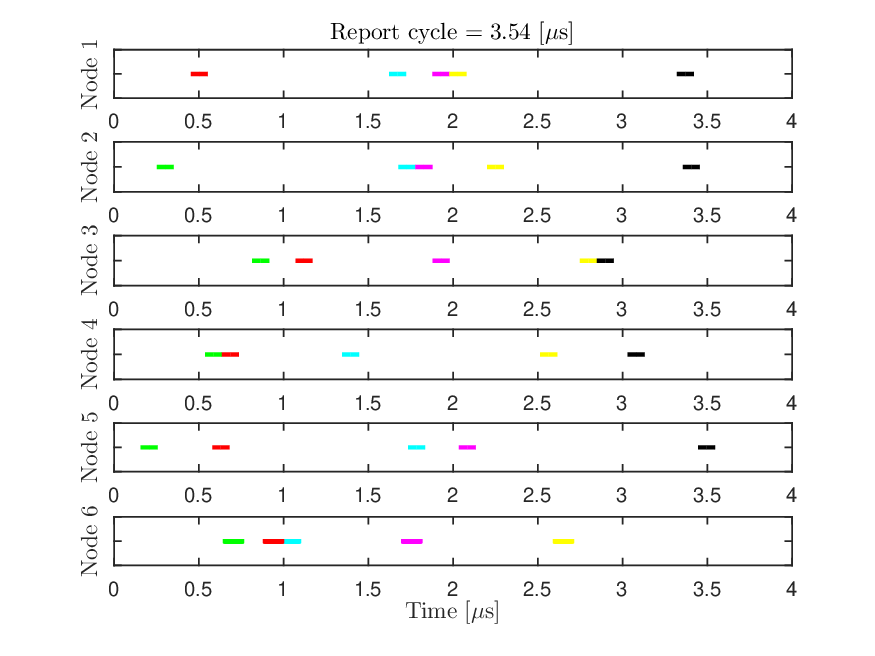}
    \label{fig:delayplot_cvx_without_outliers}} \\
    \subfloat[IPA]{\includegraphics[scale=0.55]{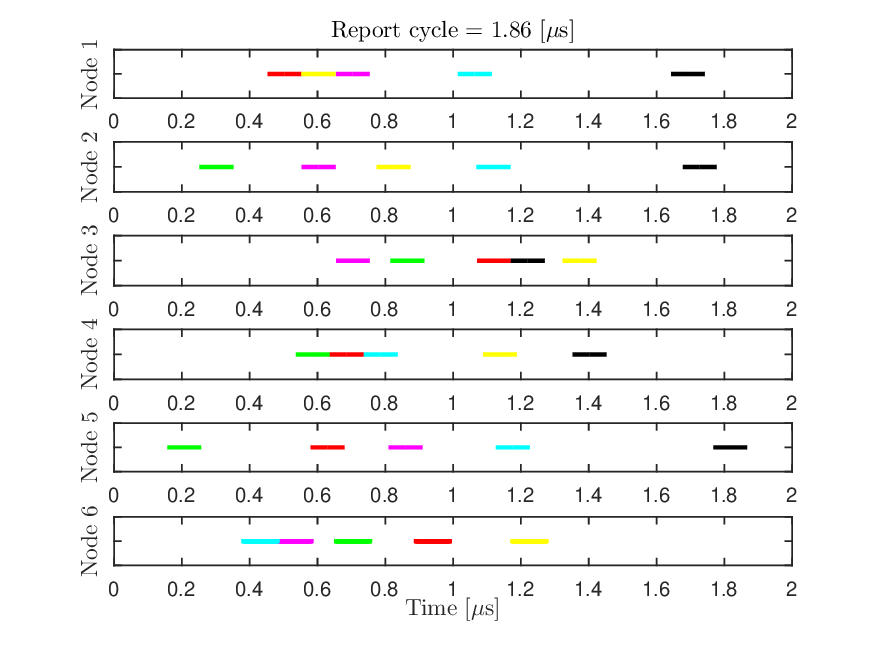}
    \label{fig:delayplot_algorithm_without_outliers}}
    \subfloat[TSP]{\includegraphics[scale=0.55]{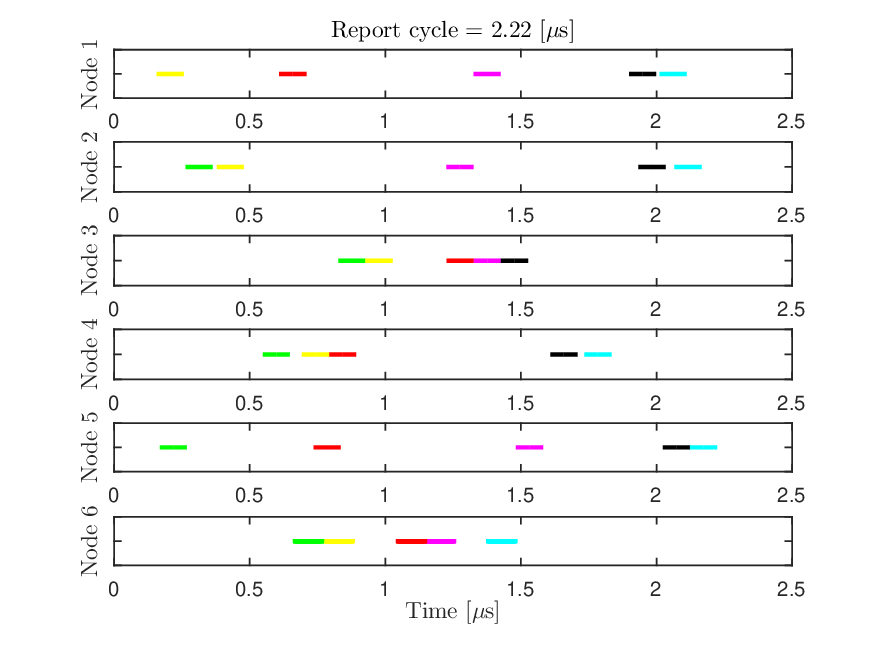}
    \label{fig:delayplot_tsp_without_outliers}}
  \caption{Network geometry with $6$ nodes scattered randomly in a 2-D plane. Received message packets at each node after introducing the computed delays from Table \ref{tab:delays-1} for network topology in Fig. \ref{fig:NetworkTopology_without_outliers} are shown for different algorithms. Notice that the message packets do not interfere. The color of the message packet is mapped to the node as shown in Table \ref{tab:colormap}}.
\label{fig:delayplot-1}
\end{figure*}

\section{Comparison with CDMA systems}
\label{sec:cdma}
In CDMA based multiple access, each node $i$ is assigned a unique spreading code, $\mbs{u_i}$, such that $\mbs{u_i} \perp \mbs{u_j} ,\mb{ }\forall i \ne j$. Each sensor transmit the packets continuously by spreading the message with its code. At the receiver, each sensor node de-spreads the signal using its unique code. Thus, in principal the  report cycle can be completed in $N\tau \mb{ [s]}$.


However, this scheme is not well suited for the all-to-all broadcast scenario described in Section \ref{sec:intro}, due to the interference originating from the near-far problem of CDMA\cite{Viterbi-book}. Unlike in many CDMA systems, this problem cannot be resolved using the classical power-control feedback. To further illustrate this, consider two $3$-node networks shown in Fig~\ref{fig:PowerCDMA}. In 
Fig~\ref{fig:PowerCDMA}~(a), $d_{\mb{AB}}=d_{\mb{BC}}=d_{\mb{CA}}$ and when all the nodes transmit messages concurrently with same power level,  the  received  message signals from all the nodes are at the same power level, and orthogonality of the different spreaded signals holds. Thus, the signal belonging to different sensors in all-to-all broadcast can be de-spread with out any interference at all the nodes. In general, for an $N$ node network to have interference free communication, we need the topology to have, $d_{ij}= \xi$, where $\xi$, is some constant. This is a rare scenario and typically does not  occur in practice. Now, consider  a network, Fig~\ref{fig:PowerCDMA}(b), with $d_{\mb{BC}}>d_{\mb{CA}}>d_{\mb{AB}}$, and a parallel transmission of an all-to-all broadcast with same power level at all nodes. The sensor node $B$ receives signals from $A$ and $C$ at different power levels, and thus $B$ will face severe interference when separating the signals from $A$ and $C$. Similar situation occur for the received signal at $A$ and $C$. The feedback power control does not solve the problem for all-to-all broadcast, as adapting the power in one node to remove interference can create interference to other nodes. As the number of nodes increases, the interference free parallel transmission becomes infeasible\footnote{For some network geometries, the interference free all-to-all communication problem can be solved as an optimization problem.}. 

Another scenario where continuous transmission is not possible are in networks which require cyclical communication. Here the transmission in the current cycle of a sensor depends on the data it received from all the other sensor nodes in the previous cycle of all-to-all communication. This kind of communication is found in distributed control and distributed computation applications as discussed in \cite{Kuhn-distributed,JO-interagent,Rantakokko-2}. 

However, we can exploit the spatial-temporal aspect of the underlying channel, where the propagation delay between nodes are much longer than the  access interval.  Here each of the $N$, nodes transmit concurrently for a duration of $\tau$, once every $\TD$ seconds. Due to the random topology, and large propagation time, the arrived pulses are spread out in time, thereby reducing the interference. It can be shown that  if a spreading code of length, $M \ge N$ is used and the shared common channel can support a bitrate of $R_b~\mb{[bps]}$, the effective bitrate per sensor, $R_s$, is given by
\begin{equation}
  \label{eq:r-cdma}
  R_s^{\mb{\tiny CDMA}}\le\frac{R_b\tau}{M\TD}.
\end{equation}

For the proposed algorithms in the paper, in each report cycle, $\TR$, each node in the network will get to transmit a message packet once for the duration, $\tau$, seconds. Therefore the effective throughput per sensor  can be computed as
\begin{equation}
  \label{eq:r-proposed}
  R_s=\frac{R_b\tau}{\TR}
\end{equation}

In the Section \ref{sec:SS}, we demonstrate in simulation the effective bitrate per sensor, $R_s$, as a function of the number of nodes, $N$, to show how the position information exploited in the proposed algorithms offer better performance compared to a CDMA based approach.

\begin{table}[!t]
  \caption{Computed delay values from proposed methods for the geometric formation defined in Fig. \ref{fig:NetworkTopology_without_outliers}.}
  \centering
  \begin{tabular}{|c|l|l|l|}
\hline
       & \multicolumn{3}{|c|}{Delay Values ($\Delta_i$s) [ns]} \\
            \cline{2-4}
     Node      & CA  & IPA & TSP\\
\hline
   $1$ &   $0$  &  $0$ &  $11.9$ \\
   $2$ &  $200.9$  & $200.8$  &  $356.8$ \\	
   $3$ &  $807.8$  & $199.2$ &  $1196.3$ \\
   $4$ &  $1341.7$ & $117.4$ &  $789.0$ \\
   $5$ &  $1821.3$ & $395.7$ &  $0$ \\	
   $6$ &  $2665.6$ & $987.8$ &  $1243.8$ \\
\hline
  \end{tabular}
  \label{tab:delays-1}
\end{table}

\begin{table}[!t]
  \caption{Mapping of colors  to nodes in   Fig. \ref{fig:delayplot-1} and Fig. \ref{fig:delayplot-2}.}
  \centering
  \begin{tabular}{|c|l|}
    \hline
    Message packet &  \\
    from node & Color \\ \hline
     $1$ & Green \\
     $2$ & Red \\
     $3$ & Cyan \\
     $4$ & Magenta \\
     $5$ & Yellow \\
     $6$ & Black \\ \hline
  \end{tabular}
  \label{tab:colormap}
\end{table}

\section{Simulation Study}
\label {sec:SS}
In the beginning of Section \ref{subsec:orthogonal}, we mentioned that we can orthogonalize the message packets by separating consecutive transmissions by a time interval equal to the maximum path delay in the network. For  the configuration in Table~\ref{tab:cfg}, the report cycle, $T_{\MS{R}}$, can be computed as below.
\begin{eqnarray}
\TD& = & \frac{\max(d_{\MS{AB}},d_{\MS{BC}},d_{\MS{CA}})}{\mu} +\tau\mb{ = 470 ns }.\\
T_{\MS{R}}&=&N \cdot \TD \mb{  =   1410 ns}
\label{eqn:val}.
\end{eqnarray}
However, if the path difference between nodes in the network topology satisfies  \eqref{eq:con},  then all the nodes can concurrently transmit; thus one report cycle can be completed in the time duration equal to the maximum path delay in the network plus the packet length, that is $470\,\mb{ns}$.
More often \eqref{eq:con} is not met. Under these circumstances we can minimize the report cycle by solving \eqref{eq:opt}.  We modified the problem so that it can be casted as a convex optimization problem. For the configuration in Table~\ref{tab:cfg}, we showed that, $(\Delta_{\MS{A}}=0\,\mb{ns, }\Delta_{\MS{B}}=84\,\mb{ns, } \Delta_{\MS{C}}=150\,\mb{ns})$
solves \eqref{eq:convex}, therefore node $C$ will transmit last after a delay of $150\,\mb{ns}$ and complete the report cycle. So one report cycle for the configuration in Table~\ref{tab:cfg} is
\begin{eqnarray}
  T_{\MS{R}}&=&\max_{ij}(\Delta_i+\delta_{ji})+\tau, \\ 
&&\forall i,j \in [A,B,C]\mb{ and }i\ne j \nonumber\\
  T_{\MS{R}}&=&150 + 370 + 100 = 620\,\mb{ns}.
\end{eqnarray}
Thus, the reduction in the report cycle equals $56\%$.

\begin{figure}[t]
  \centering
  \includegraphics[width=3in]{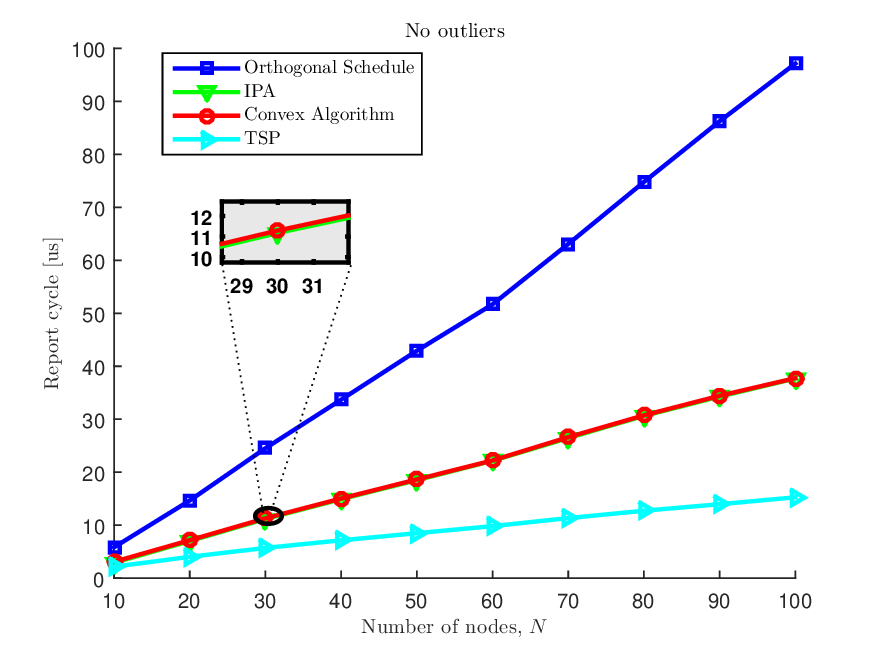}
  \caption{Performance of proposed methods as a function of $N$. Notice that as the number of nodes increases, the proposed techniques yield better performance relative to the orthogonal schedule.}
  \label{fig:Performance_no_outliers}
\end{figure}

To study the performance of the proposed methods for large networks,  we form two different formations; one with outliers, where a few sensor nodes are far apart from the rest; and another with no-outliers, where the sensor nodes are scattered uniformly. The performance is reported in terms of the time required to complete one report cycle using the proposed  methods.

\subsection{Random geometric formation  with no outliers}
\label{subsec:no-outliers}

For performance analysis with no outliers, we create a random geometric formation by scattering the nodes in a  plane. The coordinates $(x,y)$ are drawn from a Gaussian  distribution as shown below.
\begin{equation}
  \label{eq:coordinate}
  (x,y)\sim\left(\mathcal{N}(0,\sigma^2),\mathcal{N}(0,\sigma^2)\right).
\end{equation}

A typical topology of $6$ nodes with $\sigma=50~\mb{[m]}$ is shown in Fig. \ref{fig:NetworkTopology_without_outliers}. The transmission schedules for interference mitigation, using different proposed algorithms are given in Table \ref{tab:delays-1}.

\begin{figure}[!t]
  \centering
  \includegraphics[width=3in]{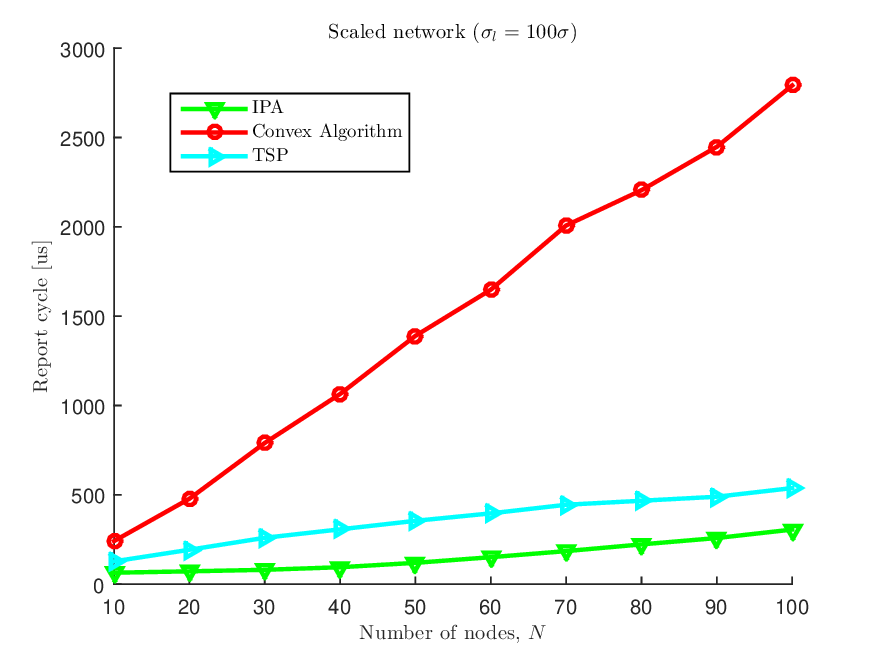}
  \caption{Performance of proposed methods as a function of $N$. Notice that as the number of nodes increases, the IPA algorithm  yield better performance relative to TSP when the network radius is larger compared to packet length.}
  \label{fig:Performance_no_outliers_scaled}
\end{figure}

\begin{figure}[!t]
  \centering
  \includegraphics[width=3in]{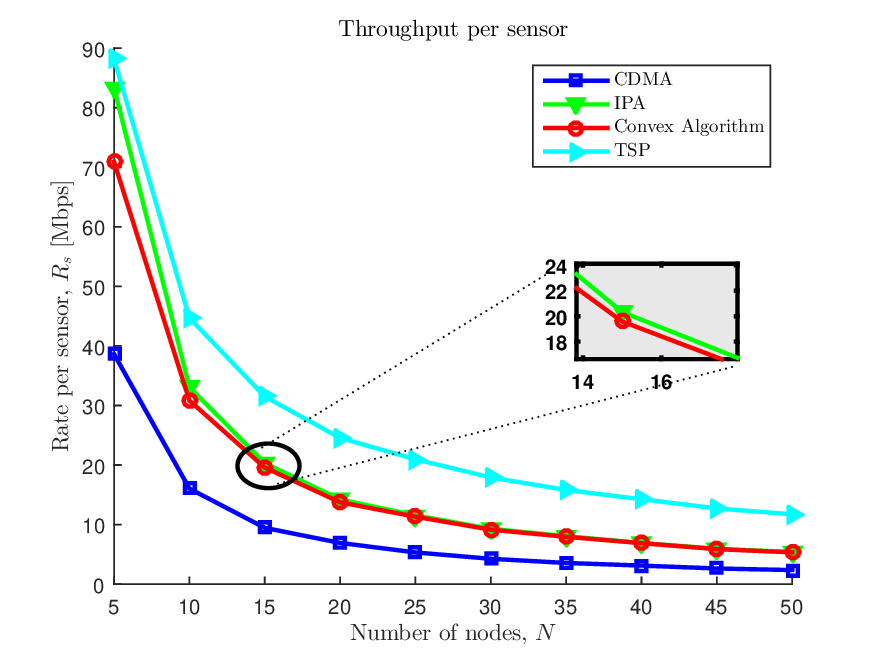}
  \caption{Bitrate per sensor, $R_s$, of proposed methods and CDMA approach as a function of $N$. Notice that the proposed methods yield better performance relative to CDMA.}

  \label{fig:Throughput_comparision}
\end{figure}

\begin{table}[!t]
  \caption{Computed delay values from proposed methods for the geometric formation defined in Fig. \ref{fig:NetworkTopology_with_outliers}.}
  \centering
  \begin{tabular}{|c|l|l|l|}
\hline
       & \multicolumn{3}{|c|}{Delay Values ($\Delta_i$s) [ns]} \\
            \cline{2-4}
     Node      & CA  & IPA & TSP\\
\hline
   $1$ & $0$ & $0$ & $401.6$ \\
   $2$ & $266.0$ & $266$ & $667.6$ \\
   $3$ & $0$ & $0$ & $0$  \\
   $4$ & $1444.7$ & $78.1$ & $1444.7$ \\
   $5$ & $939.4$ & $0$ & $939.4$ \\
   $6$ & $2528.5$ & $759.4$ & $147.7$ \\
\hline
  \end{tabular}
  \label{tab:delays-2}
\end{table}
\begin{figure*}[!t]
  \centering
    \subfloat[Network Topology]{\includegraphics[scale=0.5]{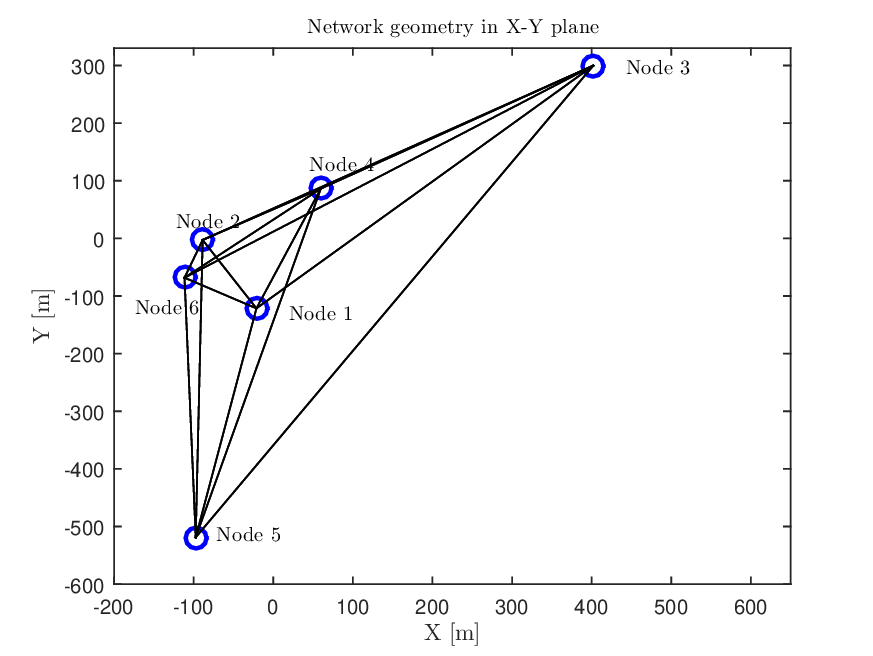}  \label{fig:NetworkTopology_with_outliers}}
    \subfloat[Convex Algorithm]{\includegraphics[scale=0.55]{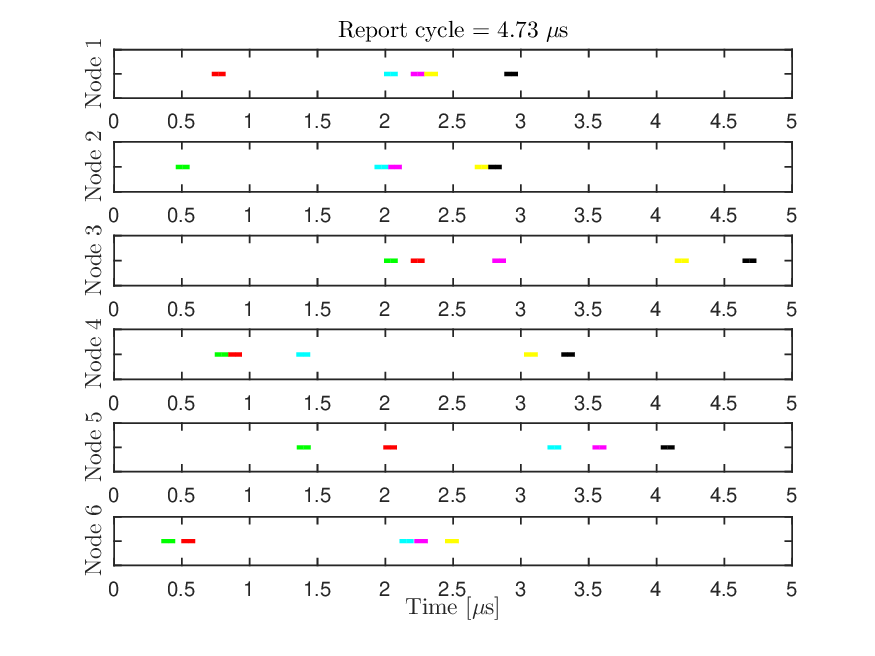}
    \label{fig:delayplot_cvx_with_outliers}} \\
    \subfloat[IPA]{\includegraphics[scale=0.55]{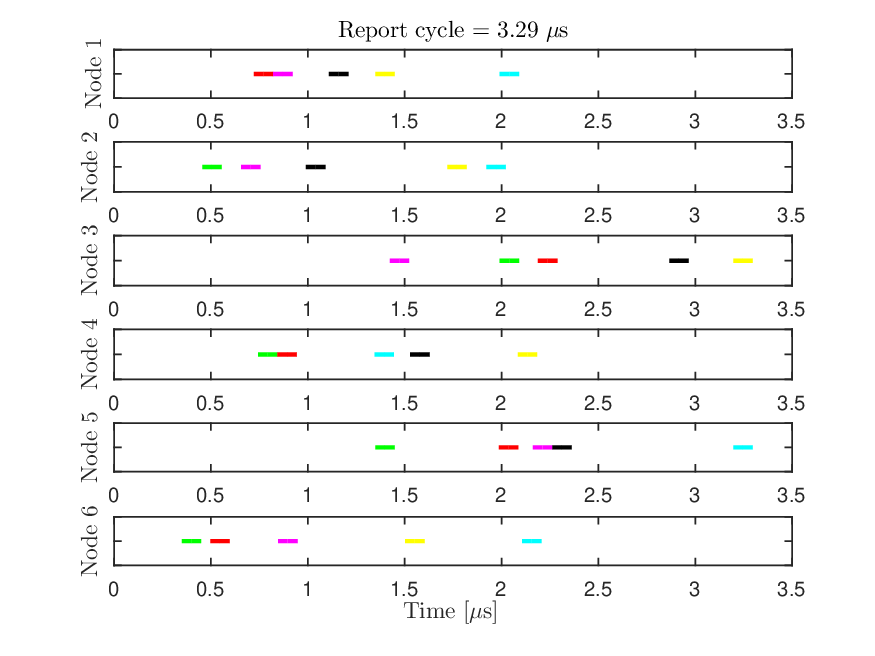}
    \label{fig:delayplot_algorithm_with_outliers}}
 \subfloat[TSP]{\includegraphics[scale=0.55]{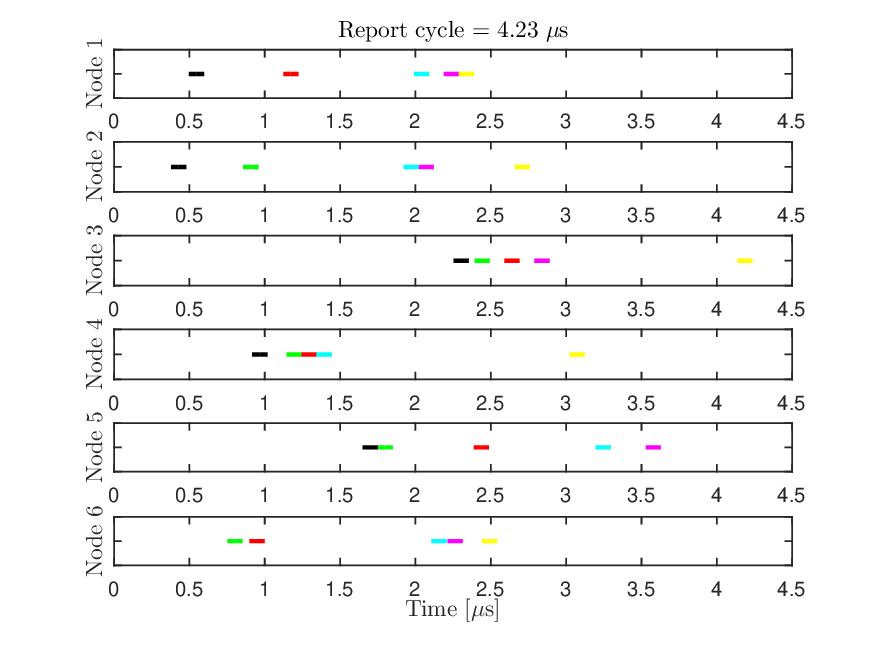}
    \label{fig:delayplot_tsp_with_outliers}}
  \caption{ Network geometry with nodes scattered randomly in a 2-D plane with outliers. Received message packets at each node after introducing the computed delays from Table \ref{tab:delays-2} for the network topology in Fig. \ref{fig:NetworkTopology_with_outliers} are shown for different algorithms. Notice that the message packets do not interfere. The color of the message packet is mapped to the node as shown in Table \ref{tab:colormap}.}
\label{fig:delayplot-2}
\end{figure*}

With these delays introduced, the received packets will not interfere with each other. The received  packets at each node are shown in Fig. \ref{fig:delayplot_cvx_without_outliers},  \ref{fig:delayplot_algorithm_without_outliers}, and \ref{fig:delayplot_tsp_without_outliers} for CA\footnote{In simulations, we employ a sequential order for CA, \emph{i.e.}, in \eqref{eq:convex-delays}, $i$, is varied from $1,\ldots,N-1$, with $\Delta_1=0$.}, IPA and  TSP algorithms proposed in this paper. Each color in Fig. \ref{fig:delayplot-1} is mapped to the messages from a specific node,  as shown in the Table \ref{tab:colormap}.

Report cycle, $\TR$, for the given set of delay values computed using CA, TSP and IPA are given by $\underset{i,j}\max(\Delta_i+\delta_{ij})+\tau$. For the example network shown in Fig. \ref{fig:NetworkTopology_without_outliers},  the report cycles are given by  $3.54\,\mu\mb{s}$, $2.22\,\mu\mb{s}$, and $1.86\,\mu\mb{s}$ for CA, TSP, and IPA respectively.

From Fig.~\ref{fig:delayplot_cvx_without_outliers} and Fig.~\ref{fig:delayplot_algorithm_without_outliers}, notice that the IPA is less constrained than the convex approach; the convex formulation requires that the order of the received message packets is the same at each receiving node. This is not the case for the IPA algorithm. This ensures tighter schedules and explains the better performance of the IPA algorithm.

To assess the performance over a large number of nodes $N$, we performed Monte-Carlo simulations. We swept the number of nodes, $N$, from $10$ to $100$ in steps of $10$ and for each $N$, $32$ distinct random geometric formations were constructed as per \eqref{eq:coordinate}. The averaged report cycle is reported in Fig. \ref{fig:Performance_no_outliers}.


Fig.~\ref{fig:Performance_no_outliers} compares the proposed algorithms to the technique of orthogonalization with scheduled transmission discussed in Section \ref{subsec:orthogonal}. Notice that for a $100$ node randomly scattered network with $\sigma=50~\mb{[m]}$, $\TR$ is reduced to approximately 1/10 for  the TSP algorithm and 1/3 for the fixed order convex algorithm and the IPA algorithm. However, if the radius of the network is scaled by a factor of $100$ by changing the variance $\sigma_l=100\sigma$, the IPA performs better than the TSP algorithm as explained in the earlier section and confirmed in simulation by  Fig.~\ref{fig:Performance_no_outliers_scaled}.

The effective rate per sensor for the CDMA approach and the proposed algorithms are as given by \eqref{eq:r-cdma}\footnote{In simulations, for \eqref{eq:r-cdma} equality is considered.} and \eqref{eq:r-proposed}. Fig.~\ref{fig:Throughput_comparision}, shows the rate per sensor for  CDMA and the proposed algorithms, assuming $R_b=1~\mb{Gbps}$ and $\tau=100~\mb{ns}$. Notice that the proposed algorithms yield better  performance in terms of bitrate per sensor, $R_s$, compared to CDMA.


\subsection{Random geometric formation  with outliers}
\label{subsec:outliers}

In this section, we will study the performance of geometric formations of the  sensor network with a few sensor nodes far apart from the rest. To create this topology, we construct $N$ nodes distributed according to a mixture of two Gaussian distributions. These distributions are as given below.
\begin{eqnarray}
  (x,y)&=&\left(\mathcal{N}(0,\sigma^2),\mathcal{N}(0,\sigma^2)\right) \label{eq:mixture1}\\
  (x_o,y_o)&=&\left(\mathcal{N}(0,\sigma_o^2),\mathcal{N}(0,\sigma_o^2)\right)
\label{eq:mixture2}
\end{eqnarray}

The node location is selected from \eqref{eq:mixture1} with probability of 2/3, and from \eqref{eq:mixture2} with probability 1/3. We set $\sigma=50$ and $\sigma_0=300$; thus for a large $N$, 1/3 of the nodes will be outliers. A typical topology with $6$ nodes is shown in Fig. \ref{fig:NetworkTopology_with_outliers}.

The transmission schedules for interference mitigation, by solving the CA, IPA, and TSP are given in Table \ref{tab:delays-2}.

With these delays introduced, the received packets will not interfere with each other. The received packets at each node are shown in Fig. \ref{fig:delayplot_cvx_with_outliers}, Fig. \ref{fig:delayplot_algorithm_with_outliers} and \ref{fig:delayplot_tsp_with_outliers} for the CA, IPA and  the TSP algorithms proposed in the paper. Each color in Fig. \ref{fig:delayplot-2} is mapped to the message from a specific node, as shown in  Table \ref{tab:colormap}.
For the example network shown in Fig. \ref{fig:NetworkTopology_with_outliers},  the report cycles are approximately given by  $4.73\,\mu\mb{s}$, $4.23\,\mu\mb{s}$, and $3.29\,\mu\mb{s}$ for the CA, TSP, and IPA  respectively.

To assess the performance over a large number of nodes $N$, we performed Monte-Carlo simulations similar to the no-outlier case with $32$ distinct random geometric formations constructed from the  mixture distribution of  \eqref{eq:mixture1} and \eqref{eq:mixture2} with probabilities of $2/3$ and $1/3$ respectively. The average report cycle is reported in Fig. \ref{fig:Performance_outliers}.

Fig.~\ref{fig:Performance_outliers} compares the proposed algorithms  to the technique of orthogonalization through scheduled transmission discussed in Section \ref{subsec:orthogonal}. Notice that for a network with $100$ nodes, $\TR$  is on average reduced to $1/8$ using TSP algorithm. This means that the net communication or update rate can be increased by a factor of  $8$  on average for the network topology with outliers. Thus, a sensor network with geometric formations having  a few outlier nodes can have higher communication rate using the proposed algorithms. The further away these outlier nodes are, the greater the benefits will be, as the algorithms can pack the information packets more efficiently there by optimally utilizing the shared common channel.

\subsection{Performance in the presence of synchronization and range errors}
With the configurations for the no-outlier topology as discussed in the Section \ref{subsec:no-outliers}, we performed Monte-Carlo simulations to assess the  average sensitivity of the algorithms to  range and synchronization errors. If there were no synchronization and range errors, then the network would have exchanged $N(N-1)$ packets of width $\tau$, without any interference, using the proposed algorithms. However, due to the errors, packets can interfere and we define  the fraction of interference free communication in the network, $F$, as
\begin{equation}
  \label{eq:F}
  F=1-\frac{\sum_i{\mathcal{I}_i}}{N(N-1)\tau}
\end{equation}
where $\mathcal{I}_i$, denotes the overlapped area of the packets at the receiving node $i$ .  The trade-off between the guard interval, $\epsilon$,  and report cycle, $\TR$, is to first pack the transmissions as closely as possible to reduce $\TR$ and then increase the guard interval, $\epsilon$, in the optimization problem, based on the environment to decrease the sensitivity to range and synchronization errors. Fig.~\ref{fig:Performance_Jitter_Range_withEpcilon}, shows the average performance of $F$, using Monte-Carlo simulations for $20$ nodes with $100$ distinct topologies constructed using the no-outlier case described earlier, with $\tau=100\,\mb{[ns]}$.


\section{Conclusion}
\label{sec:con}

\begin{figure}[!t]
  \centering
  \includegraphics[width=3in]{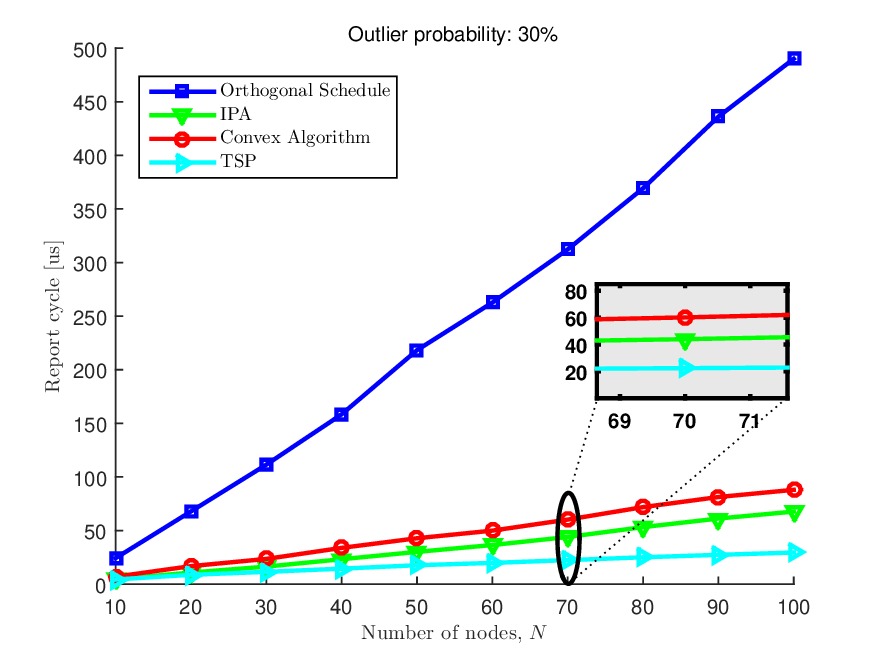}
  \caption{Performance of proposed methods as a function of $N$. Notice that as the number of nodes increases, the proposed techniques yield better performance relative to the orthogonal schedule.}
  \label{fig:Performance_outliers}
\end{figure}

\begin{figure}[!t]
  \centering
  \includegraphics[width=3in]{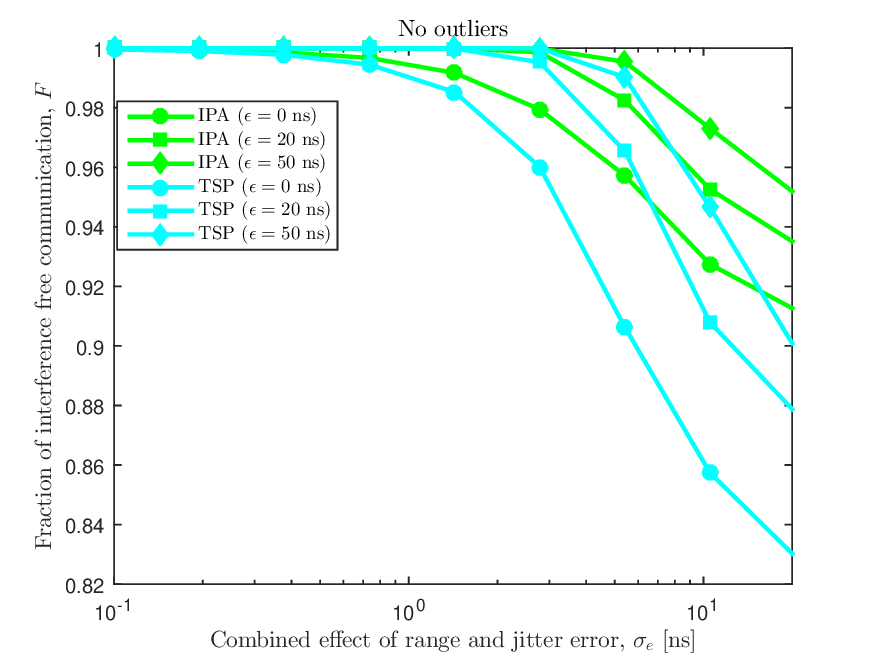}
  \caption{Interference in the system of 20 Nodes, in presence of imprecise range and clock jitter. By increasing the guard interval, $\epsilon$, by setting higher $\mathcal{P}$, as in \eqref{eq:ep}, the interference performance can be traded with report cycle. For large $\sigma_e$, the collisions are unavoidable despite the guard interval. }
  \label{fig:Performance_Jitter_Range_withEpcilon}
\end{figure}

In this paper, we proposed a methodology for utilizing the range information to arrive at  transmission schedules for high density sensor networks. Connected sensor networks need high rates of communication on a shared channel in order to have high update rates. Therefore, an optimal schedule for accessing  the shared common channel needs to be designed for efficient communication. To accomplish this, an optimization problem is formulated using range information for interference mitigation. A solution for the optimization problem is found by  CA, TSP and IPA  methods. The proposed methods are compared to the traditional time-sharing technique of separating consecutive transmissions by a time interval equal to the duration of maximum path delay in a network. The performances of the algorithms are assessed for different types of networks with varied sizes. Two different geometric formations are considered, one with a random placement of nodes with no outliers and one  with outliers. The results are demonstrated in Fig. \ref{fig:Performance_no_outliers} and Fig. \ref{fig:Performance_outliers}. A comparison with CDMA based multiple access is also presented in Fig.~\ref{fig:Throughput_comparision}. The analysis of performance degradation  due to non-idealities such as synchronization and range errors is reported in Fig.~\ref{fig:Performance_Jitter_Range_withEpcilon}. 

The three proposed algorithms performs better than CDMA or orthogonalization by scheduling one node for maximum path delay in the network. As demonstrated in  Fig.~\ref{fig:Performance_outliers}, the performance gains are higher, if the networks have few outliers in them. Each of the proposed algorithms has a clear edge over others depending on the type of the network. For example, TSP performs better than IPA and CA for general networks, however, when the network is geometrically larger in relation to the path equivalent message length, $\mathcal{L}$,  then IPA performs better than the TSP and CA as suggested by Fig.\ref{fig:Performance_no_outliers_scaled}.  IPA never performs worse than the convex algorithm. From the simulation results, it appears that the IPA will give the same solution as the CA in the worst case scenario, however, the formal proof is not known to the authors. The proposed methods assumes full connectivity, extending the methods for a partially connected network is a topic of further research.

Table \ref{tab:summary} summarizes the average complexities of the algorithms. For IPA and TSP, the worst case complexities are not known and for CA, the worst case and the average case complexities are same. Thus, IPA and TSP may not be useful in networks where real-time guarantees are needed for the schedule computations. The impact of the synchronization and range errors on the algorithms are studied. As expected, the tighter schedules are more susceptible to the interference due to the imperfect ranging and synchronization. Typical system design involves, first packing the transmissions as closely as possible to have low report cycle using the algorithms discussed and then increasing the guard interval, $\epsilon$, to decrease the interference as illustrated in Fig.~\ref{fig:Performance_Jitter_Range_withEpcilon}.
 
The performances of the proposed methods are demonstrated in simulations in order to assess the performance gains without platform or network dependencies. The in-house transceiver developed in our lab can yield very precise range information on the order of a few centimeters, as reported in \cite{Ales-2}. These transceivers could be mounted on the sensors for joint ranging and communication \cite{VJ-1,VJ-3,Ales-2}. The results from the simulations of the proposed schemes indicate that a significant improvement in performance in terms of communication rate can be achieved by using the proposed schemes. With these findings, we intend to further develop the work to implement the schemes on our in-house transceiver hardware and evaluate the performance of in-house transceiver hardware mounted sensor networks with different sizes and varied geometric formulations.

\bibliography{my}
\bibliographystyle{IEEE}

\end{document}